\documentclass[superscriptaddress,aps,preprint,pre,longbibliography]{revtex4-2}
\usepackage{graphicx} 
\usepackage{amsmath}
\usepackage{amssymb}
\usepackage[utf8]{inputenc}
\usepackage{booktabs}
\usepackage{enumitem}
\usepackage{xcolor}
\begin{document}
%\linenumbers
\title{Choosing observables that capture critical slowing down before tipping points: A Fokker-Planck operator approach}
\author{Johannes Lohmann}
\email{johannes.lohmann@nbi.ku.dk}
\affiliation{Physics of Ice, Climate and Earth, Niels Bohr Institute, University of Copenhagen, Denmark}
\author{Georg A. Gottwald}
\affiliation{School of Mathematics and Statistics, University of Sydney, NSW 2006, Australia}

\begin{abstract}
Tipping points (TP) are abrupt transitions between metastable states in complex systems, 
most often described by a bifurcation or crisis of a multistable system induced by a slowly changing control parameter. An avenue for predicting TPs in real-world systems is critical slowing down (CSD), which is a decrease in the relaxation rate after perturbations prior to a TP that can be measured by statistical early warning signals (EWS) in the autocovariance of observational time series. 
In high-dimensional systems, we cannot expect a priori chosen scalar observables to show significant EWS, and some may even show an opposite signal. Thus, to avoid false negative or positive early warnings, it is desirable to monitor fluctuations only in observables that are designed to capture CSD. Here we propose that a natural observable for this purpose can be obtained by a data-driven approximation of the first non-trivial eigenfunction of the backward Fokker-Planck (or Kolmogorov) operator, using the diffusion map algorithm. 
\end{abstract}

\maketitle

\section{Introduction}

We consider how critical transitions in stochastically forced complex systems may be anticipated by measuring increases in amplitude and temporal correlation of fluctuations in certain observables as early-warning signals (EWS). 
The complex systems of interest are a general class of heterogeneous systems with many interacting agents or scales, arising for instance in ecology, biology, social science, and the Earth system \cite{LAD13}. They are often modeled by a first-order stochastic differential equation with a non-linear, deterministic {\it drift} giving rise to possibly chaotic dynamics, and a noise process that represents unresolved scales and random disturbances by the environment, as well as control parameters that modify the drift and represent slow changes in external boundary conditions. 
A critical transition occurs when upon a parameter change a base state, i.e., a stable invariant set of the drift, loses stability and the system undergoes an abrupt transition to an alternative state. This is usually due to a collision of the base state with an {\it edge} state, which is an unstable invariant set of the drift. The stable manifold of the edge state is the basin boundary separating the base state from the alternative state. The simplest case of such a transition is a noisy saddle-node bifurcation (SNB) \cite{KUE11}, which is often considered the archetype of a tipping point (TP). %Citation?

EWS arise due to so-called critical slowing down (CSD) \cite{HAK80,NOR81,WIS84,CRO88,KLE03,HEL04,MOR24b}, which is most easily understood for the SNB. Here, as a control parameter $\mu$ crosses a critical value (e.g. $\mu_c = 0$), the leading eigenvalue of the Jacobian describing the linearized dynamics around the base fixed point crosses the imaginary axis. 
Leading up to this, for $0 <  \mu \ll 1$, the dynamics along one degree of freedom (d.o.f) becomes much slower compared to all others, and after a short relaxation time the system is confined to an (extended) center manifold, or a neighborhood thereof due to the noise forcing \cite{KNO83}. 
The drift on the manifold is one-dimensional and given by the SNB normal form after a suitable coordinate transformation. As $\mu \to 0$, CSD refers to the slowing down of the relaxation dynamics towards the equilibrium along the center manifold after an arbitrary perturbation. 

In more general cases, where the base state is a limit cycle or chaotic attractor, we also expect a decrease of the relaxation rate back to steady state after a perturbation of the system. This is plausible since upon control parameter change the underlying deterministic dynamics experience a continuous change from being stable to neutrally stable in one d.o.f, before finally becoming unstable. 
%NOTE: can we moreover motivate it should ONE d.o.f?
%continuity of the vector field under parameter change?
%mention saddle-node of limit cycles?
This generic feature makes the detection of loss of resilience to perturbations the primary avenue for predicting TPs \cite{SCH09}. 

For large-scale systems in the real world, controlled perturbations are difficult to find. But there is a permanent influence of random disturbances from the environment. 
%which may be assumed stationary in distribution. 
Such noise-driven, natural fluctuations of the unperturbed system allow one to infer the 
system's response to perturbations if linear response theory guarantees a fluctuation-dissipation theorem \cite{KUB66, HAI10}. 
The size and correlation of the fluctuations are expected to grow in tandem with the system's slowing response as the critical transition is approached, thereby forming statistical EWS. This is the other side of the coin of CSD. Growing fluctuations towards the basin boundary imply a flattening of the quasipotential \cite{Graham1991}. This happens in the direction of a particular d.o.f that is related to the location of the edge state, since the latter usually lies on the most probable path of a noise-induced escape \cite{LOH25}. 

Real-world observations have been analyzed for CSD by measuring statistical EWS of presumed critical transitions, including financial crises \cite{DIK19}, depression \cite{LEE14,WIC16}, neuron spiking \cite{MEI15}, and climate tipping points, such as the Greenland ice sheet \cite{BOE21}, Amazon forest \cite{BOU22} and Atlantic Meridional Overturning Circulation (AMOC) \cite{BOE21a, MIC22, DIT23}. But statistical false positives and false negatives can occur. The destabilization of the system in a single (critical) d.o.f implies that the increase in noise-driven variability occurs also predominantly in a single d.o.f, which gives rise to a scalar observable where the variance is expected to diverge and the autocorrelation to approach 1 at the bifurcation. 
Thus, EWS can be masked if measurements have been taken from a dynamical observable that does not sufficiently project on the critical d.o.f \cite{KUE11,BOE13,KUE13,MOR23,LOH25}. In the simple case of the SNB, this means that the observable does not follow the SNB normal form to any good approximation and hence is not subjected to significant CSD. 

Consequently, the central question that will be addressed here is what observables should be used to detect CSD. This depends on how the system under the influence of noise responds to perturbations away from its steady state, and how this response changes as a control parameter approaches its critical value. 
This can be understood in terms of the Fokker-Planck (FP) equation, which governs the temporal evolution of the probability density of the state in phase space. The density can be written as an expansion in the eigenfunctions $\psi_n$ of the FP operator $\mathcal{L}$. For a fixed control parameter any initial density will converge to the unique stationary density $\pi(\mathbf{x})$, which is the first eigenfunction $\psi_0$ with eigenvalue $\lambda_0 = 0$. The system is then in statistical equilibrium, where the contributions of all other $\psi_n$ with $|\lambda_n|>0$ have decayed. The first few $\psi_n$ (with $|\lambda_n|$ closest to 0, and assuming a discrete spectrum) signify locations in phase space where fluctuations tend to linger on finite, but long-term time horizons. 

We consider systems with a possible TP, i.e., the deterministic drift term in the governing equations gives rise to (at least) bistability, with a base and an alternative attractor. 
%NOTE: in theory also multiplicative noise can give metastability, with monostable drift...
%BUT this is normally not the scenario of B-tipping. 
A low noise strength is assumed, meaning it is small compared to the (quasi-)potential barrier height between competing attractors of the drift. It is assumed to hold for most values of $\mu$, except when very close to the bifurcation where the barrier height goes to zero. This is required for the paradigm of bifurcation-induced tipping \cite{ASH12} to be meaningful and for the concept of EWS to be useful. 
Hence, when far from the bifurcation, the system spends very long periods of time in distinct regions around the attractors, referred to as metastable states. 
Transitions between the states can be ignored, as they occur very rarely on time scales of $\mathcal{O}(1/|\lambda_1|)$ with $|\lambda_1| \ll 1$. 
While part of the invariant density $\pi(\mathbf{x})$ occupies the alternative metastable state, we hence assume that on a finite time horizon the system is in a quasi-stationary distribution concentrated entirely around the base state, where the contribution of $\psi_1$, which signifies a very slow transfer of density between the metastable states, has not decayed. 
Due to CSD, the relaxation towards the base state within the quasipotential well of the base state becomes slower along a particular mode. When close enough to the TP, this mode becomes the slowest in the system and will be expressed by the next eigenfunction $\psi_2$. 

An observable that naturally expresses increases in fluctuations related to this critical mode is proposed here to be given by the corresponding eigenfunction $\phi_2$ of the backward (adjoint) operator $\mathcal{L}^*$, also known as the generator.
$\mathcal{L}^*$ governs the temporal evolution of expectation values of observables as a function of initial states (e.g. states after a perturbation), and the first few $\phi_n$ can be interpreted as patterns of initial conditions with slowest decay towards $\pi(\mathbf{x})$. This proposition is in agreement with the framework of optimal fingerprints presented in \cite{LUC24}. 

To obtain $\mathcal{L}^*$ from data we propose to use the diffusion map (DM) algorithm \cite{COI05,COI06,NAD06}. DM has been successfully used to define generalized collective coordinates that capture the effective dynamics of complex systems \cite{NAD06,BittracherEtAl18,LueckeEtAl25}. 
It gives an approximation (discretized on the set of data points) of $\mathcal{L}^*$ induced by a stochastic differential equation with drift $\nabla \ln [\pi(x)]$, i.e., a gradient system related to the quasipotential of the underlying stochastic dynamic system \cite{Graham1991, MAR21}. %%% see e.g. Berry et al PRE 2015.
%%% write formula for quasipotential?
We show here that it yields physical observables carrying excellent EWS, also for non-gradient systems including a high-dimensional global ocean model exhibiting a TP of the AMOC. It preserves the flattening of the quasipotential (in a particular critical d.o.f) as a key property of CSD, which is not affected by non-gradient terms of the drift. Nevertheless, methods to approximate $\mathcal{L}^*$ for non-gradient systems 
can be applied, such as the extension of DM for general diffusions with multiplicative noise \cite{BAN20}, which is also implemented here and shown to give very similar results (App.~\ref{AppA}). 

%%% further, one can hope to actually see a qualitative change in the dynamics on the reduced space as a result of increasing time scale separation, which strengthens inference of CSD based on increases in variance alone. 

%%% NOTE: Include some of this? MONOTONICITY??
%becomes the first subdominant eigenfunction of $\mathcal{L}^*$ in a system constrained to the basin of attraction of the base state. 
%It is a function that varies monotonically along relaxation paths from the edge state towards the base state, and it thereby defines an observable that measures the size of fluctuations towards the edge state. 

%STRUCTURE OF THE PAPER. 
The paper is structured as follows. Sec.~\ref{sec:fokker_planck} reviews some fundamentals of FP operators and introduces notation, as well the DM algorithm to estimate the eigenfunctions of $\mathcal{L}^*$ from data. In Sec.~\ref{sec:fp_1d} and \ref{sec:fp_2d} we motivate the usage of the backward FP eigenfunctions for the purposes of EWS with simple double well potential systems in one and two dimensions. 
In Sec.~\ref{sec:diffmap} we show with conceptual models that the reconstruction of the eigenfunctions with DM indeed yields observables that carry strong EWS.  
Further, in Sec.~\ref{sec:extrapolation} we show that such observables are strictly necessary if one wants extrapolate from increasing fluctuations to forecast the timing of a TP. In Sec.~\ref{sec:veros} we apply our method successfully to a high-dimensional model of the global ocean circulation, and conclude with a discussion in Sec.~\ref{sec:discussion}. 

\section{Fokker-Planck eigenfunctions and diffusion maps}
\label{sec:fokker_planck}

Consider the $d$-dimensional state $\mathbf{X} = (X^1, ..., X^d)$, governed by an Ito diffusion equation with time-independent coefficients written component-wise as
\begin{equation}
\label{eq:system}
dX^{\gamma} = b^{\gamma} (\mathbf{X}) dt + \sqrt{\epsilon} \sigma_{\nu}^{\gamma}(\mathbf{X}) dW^{\nu},
\end{equation}
with drift $b^{\gamma}(\mathbf{X})$ and diffusion $\sigma_{\nu}^{\gamma}(\mathbf{X})$. 
%NOTE consider just using additive noise for simplicity. 
% Overdamped Langevin equation. 
The transition probability density $P(\mathbf{X}(t) = \mathbf{x} | \mathbf{X}(0) = \mathbf{x_0}) \equiv P(\mathbf{x},t | \mathbf{x_0})$ is governed by the Fokker-Planck (FP) equation
\begin{equation}
\partial_t P(\mathbf{x},t | \mathbf{x_0}) = \mathcal{L} P(\mathbf{x},t | \mathbf{x_0})
\end{equation}
with FP operator 
%%% NOTE: Is this really correctly written? First derivative not on density, but second derivate?
\begin{equation}
 \mathcal{L} (x) = - \frac{\partial}{\partial x^{\gamma}} b^{\gamma} (\mathbf{x}) + \frac{\epsilon}{2} \frac{\partial^2}{\partial x^{\gamma} \partial x^{\nu}} A^{\gamma \nu} (\mathbf{x})
\end{equation}
and diffusion tensor $A^{\gamma \nu} (\mathbf{x}) = \sigma_{\lambda}^{\gamma}(\mathbf{x}) \sigma_{\sigma}^{\nu}(\mathbf{x}) \delta^{\lambda \sigma}$. 
%Solutions are formally written as 
%\begin{equation}
%P(\mathbf{x},\tau | \mathbf{x_0}) = e^ {\mathcal{L} \tau} P(\mathbf{x_0}) = \mathcal{T}_\tau P(\mathbf{x_0})
%\end{equation}
%with transfer operator $\mathcal{T}_\tau$. 
The stationary distribution $\pi(\mathbf{x}) \equiv \lim_{t \to \infty} P(\mathbf{x},t | \mathbf{x_0})$ is an eigenfunction of $\mathcal{L}$ with eigenvalue $\lambda_0 = 0$, satisfying $\mathcal{L} \pi = 0$. The eigenfunctions of the FP operator
\begin{equation}
\mathcal{L} \psi_n(\mathbf{x}) =  \lambda_n \psi_n(\mathbf{x})
\end{equation}
have eigenvalues with $|\lambda_0| = 0 < |\lambda_1| \le |\lambda_2| \le ... < \infty$ and negative real parts. 
%then form an orthonormal basis under an inner product weighted by $\pi(\mathbf{x})$. 
We restrict our attention to a discrete FP operator spectrum. For certain sufficiently chaotic, deterministic dynamics $\mathcal{L}$ is quasi-compact in appropriate function spaces, which implies that the dominant eigenvalues are discrete and well separated from the essential spectrum \cite{EngelNagel}. 
Further, sufficiently smooth stochastic dynamical systems have compact operators provided the domain is compact \cite{Hairer11}, or in unbounded domains given certain conditions on the drift \cite{MetafuneEtAl02}. We only consider stochastic dynamical systems, since otherwise the concept of EWS is not meaningful. This can also be motivated physically in terms of the Hasselmann program \cite{Hasselmann76}, where stochastic dynamical systems are viewed as the limiting equations for macroscopic slow processes in high-dimensional deterministic systems. In particular, the interaction between chaotic multi-scale microscopic constituents in deterministic high-dimensional systems leads to an emergent stochastic behaviour at the slow macroscopic level (see, for example, \cite{GottwaldEtAl17}). A bounded domain may be physically justified too, since for practical purposes in reasonable time horizons the dynamics should be sufficiently contained to access only a bounded region.

Time-varying solutions of the FP equation can be written in the eigenfunction basis as
\begin{equation}
P(\mathbf{x},t) = \sum_{n=0}^{\infty} c_n \psi_n (\mathbf{x}) e^{\lambda_n t} ,
\end{equation}
with $c_n = \int \psi_n(\mathbf{x}) \pi^{-1}(\mathbf{x}) \pi_0(\mathbf{x}) dx$, where $\pi_0(\mathbf{x}) = P(\mathbf{x}, t=0)$. 
Since $|\lambda_0| = 0 < |\lambda_1| \le |\lambda_2| \le ...$ with nonpositive real parts, eigenfunctions with small indices decay slowest. In view of studying TPs, we consider here dynamical systems that spend a very long time in one part of phase space (a metastable set) before exhibiting a rare transition to another, and so on. In the simplest case there are two metastable sets, corresponding to neighborhoods of the attractors of the underlying deterministic dynamics, which is reflected in the spectrum of $\mathcal{L}$ as $|\lambda_1 | \ll 1$ and a spectral gap with $|\lambda_2-\lambda_1|\gg |\lambda_1|$. 
To study the dynamics before the TP, it is sufficient to consider this bistable case, since a critical transition generically consists of the collision of only a single base state and a single boundary. 

%%% BACKWARD EQUATION
The adjoint of the FP operator is the generator 
\begin{equation}
 \mathcal{L}^* (x) = a^{\gamma} (\mathbf{x}) \frac{\partial}{\partial x^{\gamma}}  + \frac{\epsilon}{2} b^{\gamma \nu} (\mathbf{x}) \frac{\partial^2}{\partial x^{\gamma} \partial x^{\nu}}. 
\end{equation}
It governs the backward Kolmogorov equation
\begin{equation}
- \partial_s u(\mathbf{x},s) = \mathcal{L}^* u(\mathbf{x},s), 
\end{equation}
which is defined on the time interval $s \in [0, T]$ for functions $u(\mathbf{x}, s) = \mathbb{E}_{x,s} [f(X_T)] \equiv \mathbb{E}[f(X_T)| X_s = \mathbf{x}]$, i.e., conditional expectation values of observables $f(\mathbf{x}, s)$ (initialized at $\mathbf{x}$), and with the final condition $u(\mathbf{x}, T) = f(\mathbf{x})$.
Employing the transformation $t = T-s$ the equation can be formulated as an initial value problem
\begin{equation}
\label{eq:bK}
\partial_t u(\mathbf{x},t) = \mathcal{L}^* u(\mathbf{x},t), 
\end{equation}
with initial condition $u(\mathbf{x}, 0) = f(\mathbf{x})$. Here, $u$ is the conditional expectation 
$u(\mathbf{x}, t) = \mathbb{E}[f(X_t)| X_0 = \mathbf{x}]$ with initial position $\mathbf{x}$. 
The solution of (\ref{eq:bK}) can be expressed as an eigenfunction expansion 
\begin{equation}
u(\mathbf{x},t) = \sum_{n=0}^{\infty} d_n \phi_n (\mathbf{x}) e^{\lambda_n t} ,
\end{equation}
with $d_n = \int \phi_n(\mathbf{x}) f(\mathbf{x}) \pi(\mathbf{x}) d\mathbf{x}$, 
and eigenfunctions $\phi_n$ satisfying $\mathcal{L}^* \phi_n = \lambda_n \phi_n$. 
The leading eigenfunction $\phi_0 (\mathbf{x})$, corresponding to the eigenvalue $\lambda_0 = 0$, is the unique solution of $\mathcal{L}^* \phi_0 = 0$ and is given by $\phi_0 = \text{const}$. This reflects the ergodicity of the underlying system, which implies that (long-term) expectation values do not depend on the initial conditions. 

%%% NOTE: notes on Koopman?? maybe just in Discussion. 
%The corresponding adjoint of the transfer operator is the (stochastic) Koopman operator. 
%NOTE: the stochastic Koopman operator ALSO evolves expectation values, and not the observable functions itself; this would be only in the deterministic case!

Any observable $g(x)$ can be expressed by an expansion in the backward eigenfunction basis with
\begin{equation}
g(\mathbf{x}) = \sum_{n=0}^{\infty} g_n \phi_n(\mathbf{x}), 
\end{equation}
and possibly approximated by a truncation thereof. Here, $g_n = \int g(x) \phi_n(x) \pi(x) dx$. Thus, $\phi_n(\mathbf{x})$ can themselves be considered as observables, and in particular the first few $\phi_n(\mathbf{x})$ are special observables with expectation values that converge only slowly because they are non-constant functions in regions of phase space where there is a slow relaxation to equilibrium. 
%%% non-constant or even monotonic? sth stronger?
%%% In the case of time scale separation, such as when approaching a bifurcation: 
From a different point of view, the subset of leading $\phi_n(x)$ are transformations of the system from the original coordinates to reduction coordinates. The reduction is meaningful in case of time scale separation, which is expected to emerge when the deterministic drift of the system approaches a bifurcation. In particular, it can be shown that the evolution of the first $k$ eigenfunctions is approximately Markovian \cite{COI08}. In this case, the long-term evolution of the system is governed by the first $k$ backward eigenfunctions. 
% and remaining fast degrees of freedom are slaved to the slow ones. 
%%% NOTE: add some more detail perhaps. 
%%% ALSO: can be shown to be optimal reduction coordinates in several ways:
%Coifman et al 2008.
%Committor function: Weinan 2005. Solution of KBE. 

%%% NOTE: comments about gradient vs non-gradient. 
For systems obeying detailed balance, i.e.,  when the drift $b(\mathbf{X})$ is the gradient of a potential and the diffusion $\sigma (\mathbf{X})$ is independent of the position $\mathbf{X}$ (additive noise), a discrete approximation to $\phi_n(x)$ can be obtained by the diffusion map (DM) algorithm given sufficient data. 
For the majority of this work, we will use this method and show that it gives good results
in the context of EWS, even for non-gradient systems, where we thus effectively reconstruct a gradient system based on the quasipotential $V_q(\mathbf{x}) \propto \ln \pi(\mathbf{x})$ of the full system. The presence of non-gradient terms does not change $\pi(\mathbf{x})$, but it gives rise to oscillatory modes that are filtered out here. 

Importantly, a generalized DM method has been introduced that allows to obtain $\phi_n(x)$ also for non-gradient drift and position-dependent (multiplicative) noise \cite{BAN20}, if one can estimate drift and diffusion at the data points. We implemented this generalized method for the high-dimensional ocean model Veros \cite{HAE18}, showing that it gives very similar results compared to the simpler DM method (see App.~\ref{AppA}). While we do not consider examples with multiplicative noise, being able to properly account for it may be crucial in practice, as variance-based methods for EWS can fail \cite{PRO23,MOR24}. 
The advantage of using the original DM method is that it requires tuning fewer parameters, and it will be more widely applicable for EWS, since drift and diffusion matrix do not need to be estimated, which can be challenging for smaller data sets in high dimensions. 

The DM algorithm defines a weighted graph on the data points, and subsequently computes the eigenvalues and eigenvectors of a random walk on this graph. To this end we define a kernel with bandwidth $\epsilon > 0$ measuring the distance of two data points $\mathbf{x}$ and $\mathbf{y}$
\begin{equation}
\label{eq:kernel}
K(\mathbf{x}, \mathbf{y}) = \exp (- (4 \epsilon)^{-1} ||\mathbf{x} - \mathbf{y}||^2).
\end{equation}
%%% Normalize 
With this, given $N$ data points $\{ \mathbf{x}_i\}_{i=1}^{N}$, construct the $N\times N$ matrix for all pairs of data points
\begin{equation}
\label{eq:distmatrix}
\tilde{K}_{ij} = \frac{K(\mathbf{x}_i, \mathbf{x}_j)}{\sqrt{p_{\epsilon}(\mathbf{x}_i) p_{\epsilon}(\mathbf{x}_j)}},
\end{equation}
with $p_{\epsilon}(\mathbf{x}) = \sum_{j=1}^{N} K(\mathbf{x}, \mathbf{x}_j)$. 
Finally, construct the row-stochastic Markov matrix 
\begin{equation}
\label{eq:markov}
 M_{ij} = \frac{\tilde{K}_{ij}}{D_i}
\end{equation}
with $D_i = \sum_{j=1}^{N} \tilde{K}_{ij}$. The first few eigenvectors $\nu_n$ of $M$, corresponding to eigenvalues $\lambda_n^{(M)}$, define the so-called diffusion coordinates $\xi_n = \lambda_n^{(M)} \nu_n$. 
In the limit $N \to \infty$ and $\epsilon \to 0$, the operator $(M-I)/\epsilon$ converges to $\mathcal{L}^*$ (i.e. the backward FP operator) \cite{COI06,Singer06,BerryHarlim16}, with $\lambda_n^{(M)} = e^{\lambda_n}$. 
%NOTE: and xi -> phi?? or nu -> phi?
%The eigenfunctions of $\mathcal{L}^*$ are related to the ones of $\mathcal{L}$ via division by the stationary density, thus allowing one to approximate the eigenfunction of the FP operator from data. 
The Euclidean distance between data points in the DM coordinates is called the diffusion distance and measures how closely two points are connected via diffusion of the Markov chain $M$. Two points $\mathbf{x}$ and $\mathbf{y}$ may have small Euclidian distance, but large diffusion distance, which can reflect that the dynamics evolve on a lower dimensional manifold. 

To evaluate the eigenfunctions approximately at points $\mathbf{y}$ not in the given data set $\{ \mathbf{x}_i\}_{i=1}^{N}$, the so-called Nyström eigenspace interpolation can be employed
\begin{equation}
\label{eq:nystrom}
\xi_n (\mathbf{y}) = \lambda_n^{-1} \sum_{i=1}^{N} \frac{\tilde{K}(\mathbf{y}, \mathbf{x}_i)}{D(\mathbf{y})} \xi_n (\mathbf{x}_i), 
\end{equation} 
with 
\begin{equation}
\label{eq:nystrom2}
\tilde{K}(\mathbf{y}, \mathbf{x}_i) = \frac{K(\mathbf{y}, \mathbf{x}_i)}{\sqrt{p_{\epsilon}(\mathbf{y}) p_{\epsilon}(\mathbf{x}_i)}}
\end{equation}
and $D(\mathbf{y}) = \sum_{i=1}^{N} \tilde{K}(\mathbf{y}, \mathbf{x}_i)$. In (\ref{eq:nystrom}), $\xi_n (\mathbf{x}_i)$ denotes the entry of the n-th eigenvector corresponding to the i-th data point. 

In our DM implementation we normalize each data variable to have unit variance. Furthermore, we remove a small number of outliers so that we can use a smaller $\epsilon$ to obtain a better approximation of the backward eigenfunctions. 
Individual outliers are those data points that where the minimum of the distance to all other points is largest. 
We find $n=15$ points with the largest minimum distance and remove the respective columns and rows in the distance matrix $K(\mathbf{x}_i, \mathbf{x}_j)$ in (\ref{eq:distmatrix}). 
Thereafter we remove $n=10$ double outliers, which are those pairs of points where the second smallest distance to all other points is largest. 
The number of removed outliers has been chosen by trial and error to give the best performance across all data sets with typical sample size of about 10,000. If this step is skipped, often a quite large $\epsilon$ is needed to prevent the first diffusion coordinates from merely acting to cluster individual outliers against the rest of the data. 

\section{Results}

\subsection{Interpretation of Fokker-Planck eigenfunctions in one dimension}
\label{sec:fp_1d}

\begin{figure}%[floatfix]%!htb
\includegraphics[width=0.99\textwidth]{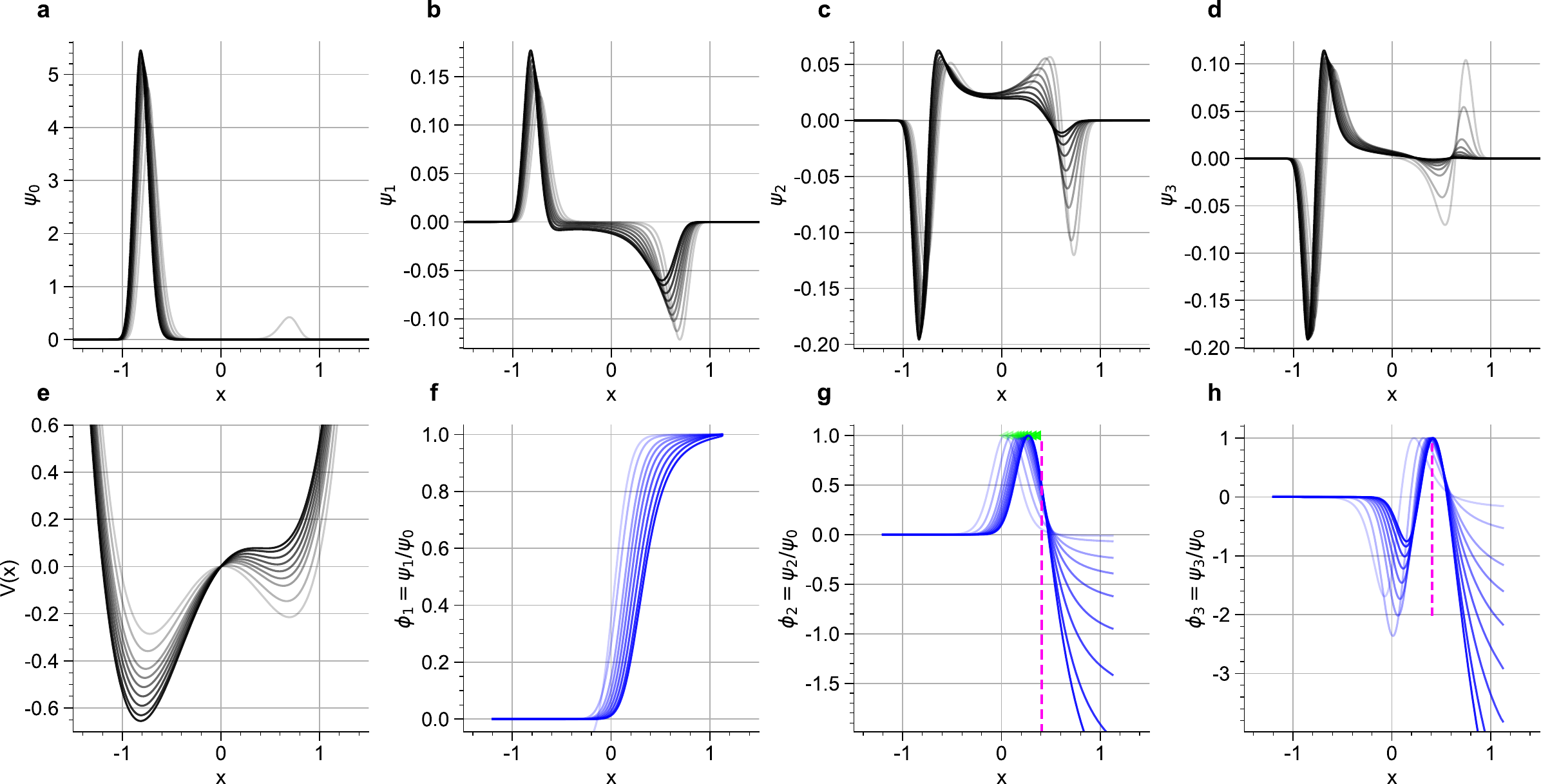}
\caption{\label{fig:eigenfunctions_dw1D} 
Eigenfunctions of the Fokker-Planck operator for the one-dimensional double-well model (\ref{eq:dw1}) for a range of parameter values from $\beta=0.05$ to $\beta=0.53$ drawn with increasing contrast. The bifurcation occurs at $\beta \approx 0.544$. 
The eigenfunctions are estimated by eigendecomposition of the discrete approximation of the FP operator via the scheme of Chang and Cooper \cite{CHA70}.
{\bf a}-{\bf d} The first four eigenfunctions of the forward FP operator for $\sigma = 0.25$. 
{\bf e} Associated potential $V(x)$. 
{\bf f}-{\bf h} The first three non-trivial eigenfunctions of the backward operator $\phi_{1,2,3}$ (rescaled for each parameter value to have a maximum value of 1). The green dots in {\bf g} indicate the locations of the saddle point for the respective parameter values. The vertical dashed line is the inflection point of the potential in the shallower well, which is independent of $\beta$. 
}
\end{figure}

We first study the FP eigenfunctions in a one-dimensional system with additive noise
\begin{equation}
\label{eq:dw1}
dX_t = -\left (\frac{d}{dx} V(X_t) \right )dt + \sigma \, dW_t,
\end{equation}
and double-well potential
\begin{equation}
V(x) = x^4-x^2+\beta x 
\end{equation}
where $\beta$ is a control parameter. The potential is shown in Fig.~\ref{fig:eigenfunctions_dw1D}e for different values of $\beta$ .
The deterministic drift of the system undergoes a saddle-node bifurcation at $\beta_c = \sqrt{8/27} \approx 0.544331$, where one of the potential wells disappears. From here on we will simply refer to this as bifurcation or TP, also when referring to the stochastic system.  We determine the forward eigenfunctions $\psi_n$ numerically by a discrete approximation of $\mathcal{L}$ using the finite difference method by Chang and Cooper \cite{CHA70}, and then an eigendecomposition of the obtained matrix using an implicitly restarted Arnoldi method (scipy.sparse.linalg.eigs package implementation of ARPACK).
Since this is a gradient system the forward and backward eigenfunctions are related by $\phi_n (\mathbf{x}) =  \pi (\mathbf{x})^{-1} \psi_n (\mathbf{x})$, with the same, real $\lambda_n$. Hence, we can determine both $\psi_n$ and $\phi_n$, which is shown in Fig.~\ref{fig:eigenfunctions_dw1D}. 

As $\beta$ is increased from zero towards the bifurcation, for low noise $\pi(\mathbf{x})$ quickly becomes heavily asymmetric with a dominant peak at the deeper potential well (Fig.~\ref{fig:eigenfunctions_dw1D}a), and a peak around the shallower well that is orders of magnitude smaller. The $\psi_n$ for small $n > 0$ correspond to distinct patterns that modify the density such that it takes longest until statistical equilibrium is reached,
given that the initial density $\pi_0(\mathbf{x})$ projects significantly on those pattern.
%NOTE: IMPROVE. 
%given it is in the support of the initial density $\rho_0$ 
%given that the initial density is not the stationary density. 

The slowest decaying pattern $\psi_1$ %is prominent, meaning c1 large
describes the situation where the probability mass in one of the wells is initially larger as it should be according to $\pi(\mathbf{x})$ (Fig.~\ref{fig:eigenfunctions_dw1D}b). For the equilibration of such a configuration, part of the probability mass needs to diffuse uphill and overcome the potential barrier. In systems with low noise this is associated with a long time scale, and $|\lambda_1|$ is approximately equal to the escape rate out of the shallow well. %say Kramers, or even give formula?
% Kramers approx. will break down though when close to bif. not low noise anymore. 

The pattern $\psi_2$ gives a large contribution when the initial density is concentrated more prominently in the vicinity of the saddle (as compared to $\pi(\mathbf{x})$).
%%% or less prominently actually; all is symmetric!
The function shows minima slightly outwards (larger $|x|$)) of the two stable fixed points, and a broad double maximum around the saddle point (Fig.~\ref{fig:eigenfunctions_dw1D}c). 
%NOTE: IMPROVE FURTHER!
% I THINK ARGUMENT IS THE SAME AS OF PSI2: it is symmetric, but there is simply 
% much more final mass close to the fixed points, thus time scale should be 
% interpreted as relaxation time down from the saddle vicinity. 
This pattern can be interpreted as additional mass that survives outside the vicinity of the two minima for some time $|\lambda_2|^{-1}$ due to the asymmetry of each well, i.e., the smaller curvature of the potential towards the saddle. 
In other words, the relaxation towards equilibrium is slower in the vicinity of the saddle and on the sides of the wells that are facing the saddle. As the bifurcation is approached, the segment of the potential within the shallow well that faces the saddle becomes more flat, and thus $\psi_2$ is the relevant mode carrying CSD. 
%NOTE: maybe also say that this is just the other side of the coin. 
%%% of course lambda1 also changes!
In the one-dimensional case, higher eigenfunctions are less important for our analysis, representing higher-order corrections (see Fig.~\ref{fig:eigenfunctions_dw1D}d for $\psi_3$). 

With these considerations on $\psi_n$ one may interpret the backward eigenfunctions $\phi_n$. As mentioned above, $\phi_0$ is constant due to the ergodicity of the system. $\phi_1$ shows a sigmoidal shape, with plateaus around the two fixed points. On time scales smaller than the mean escape time from the shallow well, observables thus have different, quasi-constant expectation values that depend on which is the starting basin. % more generally: depending on where the initial density is concentrated
% more prominently
The transition zone of the sigmoid function with its midpoint at the saddle becomes narrower for decreasing noise levels.

While the constant $\phi_0$ can be called the ``trivial'' eigenfunction, and $\phi_1$ the dominant eigenfunction since $|\lambda_2-\lambda_1|\gg |\lambda_1|$, we refer to $\phi_2$ as the first subdominant eigenfunction. $\phi_2$ peaks close to the saddle point, 
%NOTE: why not AT the saddle point...?
and converges to 0 at the deep well while reaching lower values in the shallow well. 
On time scales of order $|\lambda_2|^{-1}$, expectation values are thus altered when starting close to the saddle. 
$\phi_3$ (and similarly higher eigenfunctions) is non-monotonic within the shallow well. It first increases towards the inflection point in the shallow well (Fig.~\ref{fig:eigenfunctions_dw1D}h), and then decreases again towards the saddle. For the purpose of EWS, we are, however, interested in observables that are monotonic from the base attractor towards the edge state, because otherwise fluctuations of increasing length along the critical d.o.f towards the edge state due to CSD are suppressed in the measured observable. 
% loosely speaking, observable should be injective with respect to the critical d.o.f. 
%CSD implies that fluctuations of increasing length occur along the critical d.o.f towards the edge state. An eventual non-monotonicity of a chosen observable would not be consistent with this increase, and hence non-monotonic observables can be discarded as appropriate observables for the detection of TPs.

\begin{figure}%[floatfix]%!htb
\includegraphics[width=0.99\textwidth]{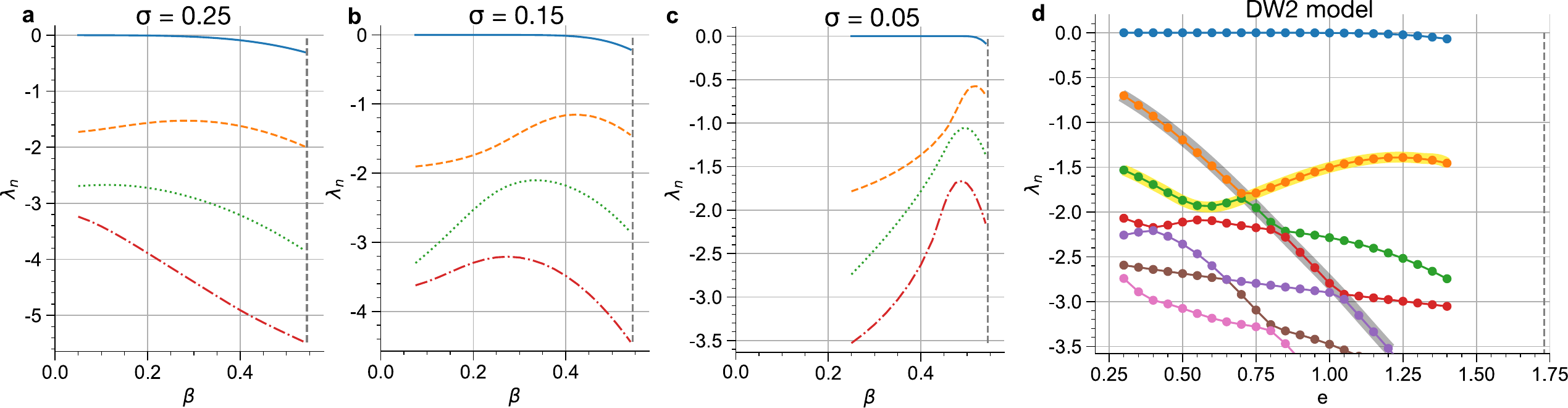}
\caption{\label{fig:eigenvalues_dw} 
{\bf a-c} Eigenvalues $\{\lambda_1, \lambda_2, \lambda_3, \lambda_4 \}$ of the FP operator for the one-dimensional double-well potential as function of the control parameter $\beta$, for different noise levels $\sigma$. The critical value corresponding to the bifurcation is marked by the vertical dashed line. 
{\bf d} Eigenvalues $\{\lambda_1, \lambda_2, ... , \lambda_7 \}$ of the FP operator of the two-dimensional double-well (\ref{eq:gradient}) as a function of the control parameter $e$, using the noise level $\sigma = 0.3$. The bifurcation point is marked with the vertical dashed line.
}
\end{figure}

%%%Eigenvalues as control param is changed. 
Depending on the noise level, the CSD as $\beta$ is changed towards the bifurcation may be reflected in the eigenvalues. Since $\psi_2$ captures the slowing of relaxation towards equilibrium as the curvature in the shallow well decreases, the relaxation rate $|\lambda_2|$ should go towards zero as a manifestation of CSD. This can indeed be seen for low noise levels in Fig.~\ref{fig:eigenvalues_dw}b,c. But for finite noise levels $|\lambda_2|$ remains finite (Fig.~\ref{fig:eigenvalues_dw}a), and even slightly increases close to the bifurcation. It is bounded by the noise-induced escape rate $|\lambda_1|$, which increases drastically and becomes O(1) at the bifurcation, at which point the potential is so flat that the relevant time scale for the decay of $\psi_2$ is not set by deterministic relaxation, but pure diffusion dynamics. Additionally, the approach of saddle and fixed point in the shallow well may play a role. 

%CITE: Cerou?
% Cerou 2013:
% for a totally flat potential, the length of a reactive path goes to infinity at 
% rate 1/eps; for a non-degenerate (curvature nowhere 0, but gradient zero at point 0)
% potential the length of a reactive path goes also to infinity, but at a slower rate,
% namely log(1/eps). Here, sigma = sqrt(2eps)
%%% NOTE: maybe even stronger statement: for finite noise the relaxation part 
% of the noisy trajectories are far from the deterministic relaxation dynamics, and 
% they do not slow down at all. 

%%% arguing pathwise: 
%In the low noise limit, the most likely trajectories of individual realizations of the stochastic processes are equal to (time-reversed) deterministic relaxation trajectories. 
%%% more explicitly: The relaxation of the density ``downhill'' from the configuration psi3 to psi1 would become infinitely slow at the bif point, if it were following the relaxation trajectories. 

\subsection{Eigenfunctions in two dimensions}
\label{sec:fp_2d}

\begin{figure}%[floatfix]%!htb
\includegraphics[width=0.95\textwidth]{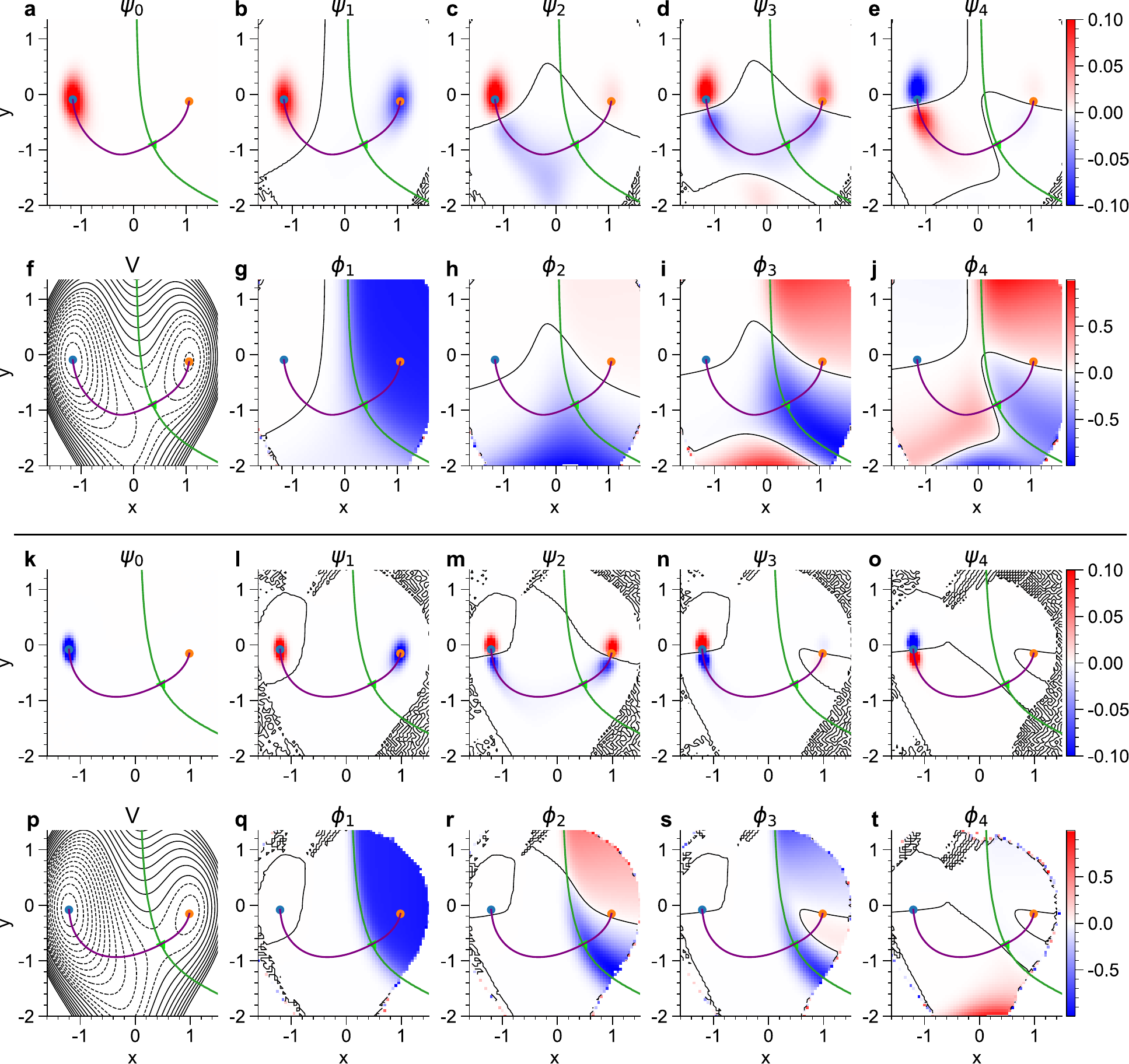}
\caption{\label{fig:grad2D_eigen} 
{\bf a-j} Forward ({\bf a-e}) and backward ({\bf g-j}) FP eigenfunctions of the two-dimensional double well model (\ref{eq:gradient}) with $\sigma = 0.6$ and control parameter $e=0.5$, computed using the method by Chang and Cooper \cite{CHA70}.
The black contour depicts the level where $\psi_n = \phi_n = 0$. 
The potential $V(x,y)$ of the system is shown as level sets in ({\bf f}). The instanton (computed by the method in \cite{KIK20}) is drawn in purple, and the basin boundary in green. 
{\bf k-t} Same but for the model with $\sigma = 0.3$ and control parameter $e=1.0$, which is closer to the bifurcation at $e\approx 1.73$ compared to the case in panels {\bf a-j}.
Note there is numerical noise due to the very low probabilities that occur at the steepest parts of the potential around the boundaries of the domain. This produces numerical artefacts in the zero contour-line of the eigenfunctions, where erroneously the values in the computed eigenfunctions rapidly alternate in sign. 
}
\end{figure}

In the previous one-dimensional example, there were no other slow d.o.f. that compete with the critical d.o.f. to be in the position of mode $\phi_2$. 
%NOTE: one could have a weird potential with flat parts, which would first be phi2. 
But generally the situation is different, especially in high-dimensional systems with multi-scale behaviour, where the correct physical mode first needs to slow down enough so that it emerges as $\phi_2$. We illustrate this with a system of two variables $x$ and $y$ in the double-well potential  
\begin{equation}
\label{eq:potential}
V(x,y) = x^2(x^2 + y^2 - a) + y\frac{cy+d}{x^2 + b} + ex. 
\end{equation}
Adding Gaussian white noise independently to both variables yields the system 
\begin{equation}
\label{eq:gradient}
\begin{pmatrix}
dx_t\\[\jot]
dy_t
\end{pmatrix}=\begin{pmatrix}
 -\frac{\partial V}{\partial x}\\[\jot]-\frac{\partial V}{\partial y}
\end{pmatrix} dt + \begin{pmatrix}
\sigma_x \, dW_{x,t}\\[\jot]\sigma_y \, dW_{y,t}
\end{pmatrix}, 
\end{equation}
where $W_{x,t}$ and $W_{y,t}$ are independent, standard Wiener processes. $a=2.5$, $b=0.5$, $c=0.2$, $d=0.5$ are chosen, and $e$ is the control parameter. For small $e$, there are two stable fixed points and one saddle point in the deterministic system. There is a saddle-node bifurcation of the deterministic drift at $e_c \approx 1.73$, where the potential well with $x>0$ disappears. Figure~\ref{fig:grad2D_eigen}f shows isolines of the potential, as well as the fixed points and basin boundary at $e=0.5$.

The first non-trivial eigenfunction  $\psi_1$ is again related to the slow transport of density from one well to the other (Fig.~\ref{fig:grad2D_eigen}b), and accordingly the backward function $\phi_1$ is approximately constant in the two basins (Fig.~\ref{fig:grad2D_eigen}g). Next, compared to the 1-d double well there is an additional eigenmode because of the slow time scale from the generally slower deterministic dynamics in the $y$ direction. 
Far from the bifurcation, $\psi_2$ represents probability mass that is more slowly contracted in the $y$-direction and for some time ($|\lambda_2|^{-1}$) has the tendency to linger at strongly negative $y$ values (Fig.~\ref{fig:grad2D_eigen}c) instead of converging to either fixed point. Hence, $\phi_2$ identifies initial conditions that take longest to converge to either of the two wells along the $y$-direction (Fig.~\ref{fig:grad2D_eigen}h).
%NOTE perhaps supp figures to emphasize this! both figure with absolute value of vector field, and eigenfunctions of symmetric 2D potential. 
The next mode $\psi_3$ corresponds (at this value of the control parameter $e$) to $\psi_2$ of the one-dimensional case, i.e., a result of slow convergence to the fixed points in the more flat parts of the asymmetric potential wells towards the saddle point (Fig.~\ref{fig:grad2D_eigen}d). Correspondingly, $\phi_3$ peaks near the saddle, and it shows that in particular initial conditions starting near the stable manifold of the saddle will lead to a slow relaxation of conditional expectation values. $\psi_4$ is a higher-order, antisymmetric pattern, analogous to $\psi_3$ of the 1-d double well.

As $e$ approaches the bifurcation, the eigenvalues of the abovementioned patterns cross (Fig.~\ref{fig:eigenvalues_dw}d), leading to a different ordering of the modes (Fig.~\ref{fig:grad2D_eigen}k-t). The eigenvalue of the pattern associated with the d.o.f. in the $y$-direction decreases (gray band in Fig.~\ref{fig:eigenvalues_dw}d), and the pattern drops to higher $n$. 
Instead, the pattern related to the low potential curvature towards the saddle (yellow band in Fig.~\ref{fig:eigenvalues_dw}d) becomes $\phi_2$ (Fig.~\ref{fig:grad2D_eigen}r), i.e., the pattern subdominant only to the pattern $\phi_1$ that reflects noise-induced escape. 
In order for the critical d.o.f. to reach the position of $\phi_2$ before a transition, it is crucial that the noise is low enough. 
If this is not the case and there is a very slow (and unchanging) d.o.f, the moment when the critical d.o.f gets slow enough, i.e., when $|\lambda_3|$ approaches $|\lambda_2|$, may only occur at a parameter value where noise-induced escape is already common, i.e., when $|\lambda_1|$ approaches $|\lambda_2|$ and $|\lambda_3|$ from below and reaches the same order of magnitude. 
%NOTE: maybe lambda1 even crosses? it could be that lambda2 so slow that we cannot speak of ``common'' noise-induced transitions

\subsection{Eigenfunction reconstruction from diffusion maps and observables for early-warning signals}
\label{sec:diffmap}

%%% reconstruction with diffusion maps.
We now reconstruct $\phi_n(x)$ via the DM approach from data of the 2-d double-well (\ref{eq:gradient}) obtained by simulation with an Euler-Maruyama scheme with time step $dt=0.005$. We simulate an ensemble of 100 uniformly distributed initial conditions covering both wells for a fixed simulation time $T=100$, allowing the ensemble to converge to $\pi(\mathbf{x})$. We only use data after $t=75$, i.e., we discard any transient dynamics. Finally, the simulated data is subsampled and all ensemble members combined to yield a set of 15,000 data points. 

\begin{figure}%[floatfix]%!htb
\includegraphics[width=0.85\textwidth]{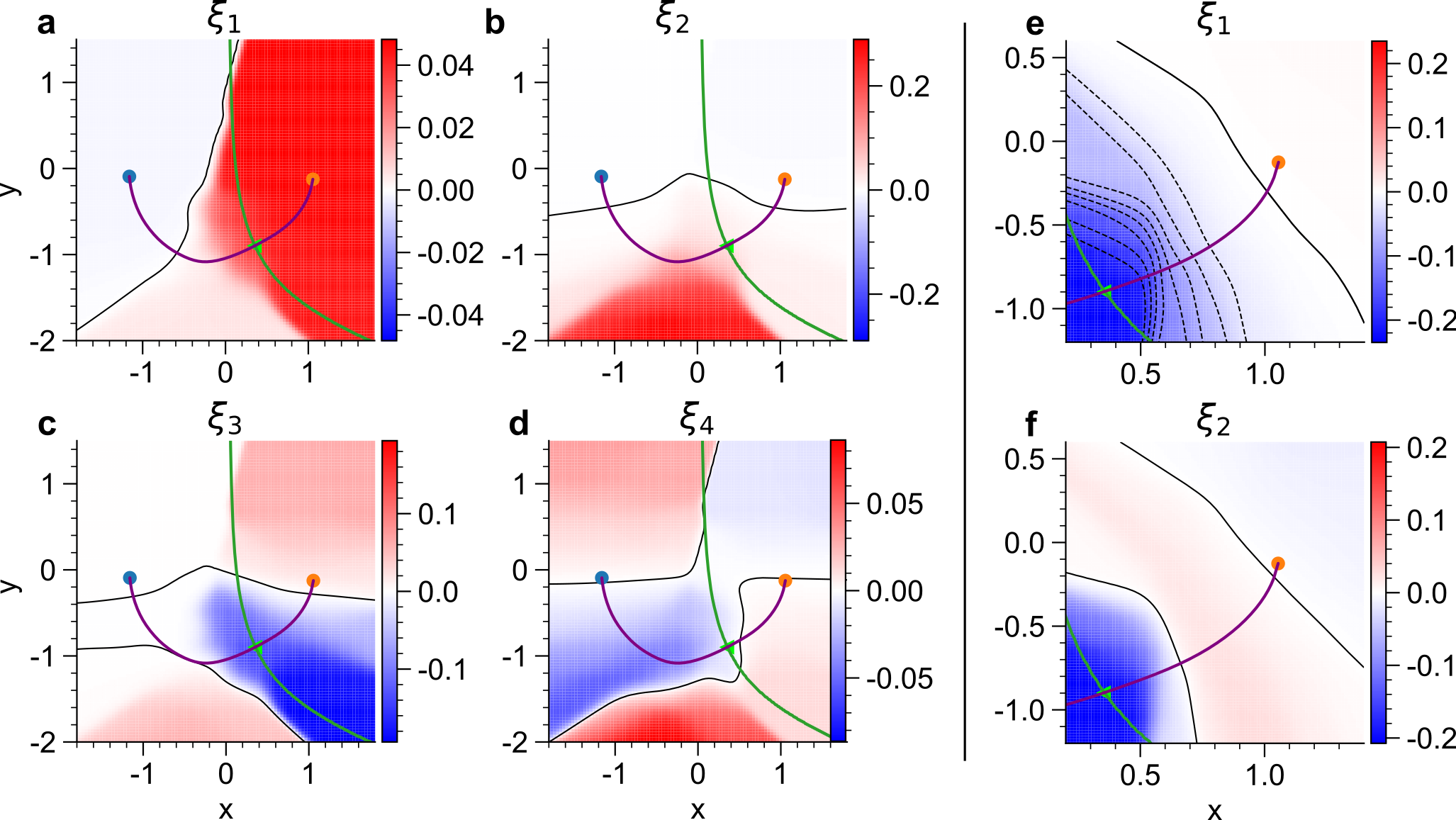}
\caption{\label{fig:grad2D_eigen_reconstr} 
Reconstruction of eigenfunctions of the 2-d double-well (\ref{eq:gradient}) using the DM algorithm. 
{\bf a-d} First four diffusion coordinates $\xi_{1,2,3,4}$ that approximate $\phi_{1,2,3,4}$, obtained at control parameter $e=0.5$ and noise strength $\sigma = 0.6$, and estimated by DM on simulated data sampling the whole phase space. 
The diffusion coordinates are evaluated at evenly spaced grid points using Eq.~\ref{eq:nystrom}-\ref{eq:nystrom2}.
{\bf e,f} First two non-trivial diffusion coordinates of (\ref{eq:gradient}) from simulated data restricted to dynamics that remains in the shallow well, with control parameter $e=0.5$ and noise strength $\sigma = 0.3$.
}
\end{figure}

The scaled eigenvectors $\xi_n$ (i.e. the diffusion coordinates) obtained from a spectral decomposition of the Markov matrix (\ref{eq:markov}) define functions that can be evaluated approximately at any point in the original phase space via Eq.~\ref{eq:nystrom}-\ref{eq:nystrom2}. 
Evaluation on an evenly spaced grid for 2-d double-well shows that the first few non-trivial diffusion coordinates $\xi_{1,2,3,4}$ are indeed in good qualitative agreement with the (scaled) eigenfunctions $\phi_{1,2,3,4}$ obtained from the discretized FP operator (compare Fig.~\ref{fig:grad2D_eigen_reconstr}a-d and Fig.~\ref{fig:grad2D_eigen}g-j). 

We now restrict our attention to the scenario of bifurcation-induced tipping, where the system resides in one of the metastable sets, and where we consider time scales much shorter than $|\lambda_1|^{-1}$ associated with noise-induced escape.  
We consider the well that contains the base state with $x>0$ (cf. Fig.~\ref{fig:grad2D_eigen_reconstr} or  Fig.~\ref{fig:grad2D_eigen}). Assuming a slowly varying control parameter, the system relaxes towards a quasi-stationary distribution $p_{qs}(x) \approx \psi_0 (x) + c_1 \psi_1(x)$ at any given instantaneous control parameter value. 
Here, $c_1 \psi_1$ compensates $\psi_0$ such that all mass is concentrated in the shallow well. The dominant eigenfunction $\phi_1$ is approximately constant in the shallow well, and thus variations in expectation values are determined by $\phi_2$ and onward. When sufficiently close to the bifurcation, the first backward eigenfunctions are almost constant $\phi_2 \approx \phi_3 \approx \phi_4 \approx 0$ in the basin of the alternative state (Fig.~\ref{fig:grad2D_eigen}q-t). 
%%% NOTE: more precise? ``the interpretation of the eigenfunctions is preserved''?
Hence, it should be possible to approximate them from data restricted to the basin of the base state.

%%% NEED connection to reaction coordinates, and maybe even COMMITTOR = optimal reaction
%%% coordinate. 
%%% CITE: Nadler 2006. NOTE: they say second eigenfunction (here phi1) is the reaction
%%% coordinate. 
%%% reaction coordinates = slow variables. this makes somewhat sense since the 
%%% fast eigenfunctions have decayed. 
%%% for EWS: contrast of fast decaying directions and the direction of CSD!!!

%%% or merely: reduction coordinates. this could then be close to center manifold
%%% as we approach the bifurcation. 

%%% OPTIMALITY of truncated diffusion coordinates. see Coifman 2008. 
%%% they also have Lemma showing that for spectral gap at k-th eigenvalue,
%%% the first k diff map coordinates are approximated Markovian, hence very good 
%%% reduction coordinates. 
%%% BUT: they also say this may not be appropriate when starting at very low density 
%%% points. We are not doing that necessarily, even though we start at shallow well. 

%%% Coifman 2008 argue that for double well the first eigenfunction as reduction 
%%% coordinate is a monotonic function of the arclength of the instanton curve of transitions in the limit of small noise. (dont think they have proof for this). 

We compute DMs from simulation data with initial conditions restricted to a square domain around the base state that lies entirely within its basin. The initial conditions quickly converge to the quasi-stationary distribution $p_{qs}(x)$ in the shallow well and the transient during equilibration is discarded. A small noise strength $\sigma=0.3$ is used, ensuring that noise-induced transitions to the other well during the simulation time are extremely rare. Realization leading to transitions are discarded. 
The features of the first two non-trivial eigenfunctions (Fig.~\ref{fig:grad2D_eigen_reconstr}e,f) are consistent with the corresponding (higher) eigenfunctions obtained from the full state space (Fig.~\ref{fig:grad2D_eigen}r,s or Fig.~\ref{fig:grad2D_eigen}i,j). Specifically, the level sets show that $\xi_1$ is increasing monotonically and non-linearly towards the saddle. Level sets of $\xi_2$ have similar shape, but do not exhibit a monotonic increase toward the saddle. 
In fact, there is a quadratic relationship between $\xi_2$ and $\xi_1$. 
%relation to center manifold? unclear: of course one dominant degree of freedom, 
% but other degrees of freedom on one hand fast decaying modes, on the other hand 
% they define the center manifold...?

\begin{figure}%[floatfix]%!htb
\includegraphics[width=0.99\textwidth]{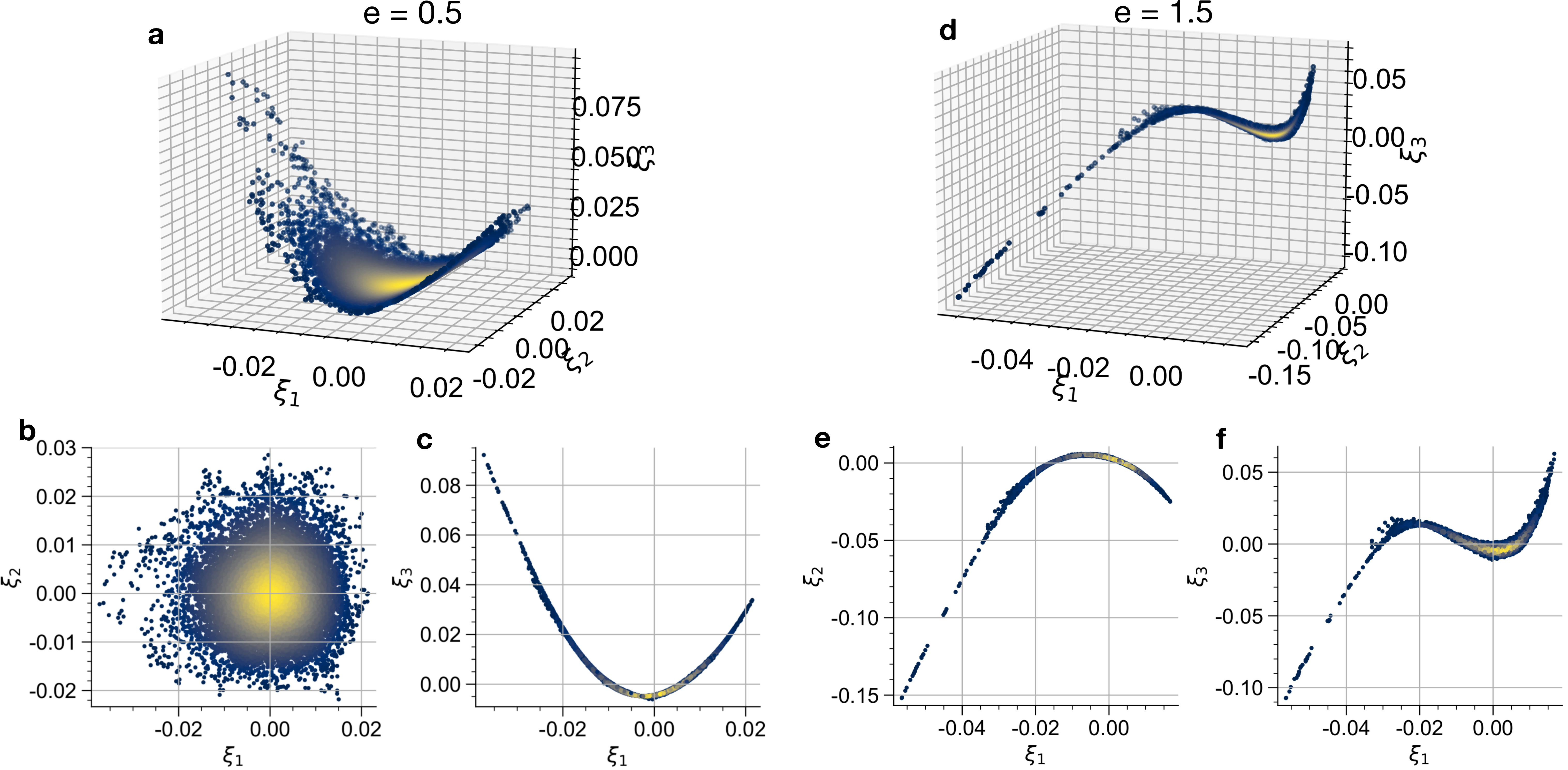}
\caption{\label{fig:grad2D_xi} 
{\bf a, d} Simulated data points of the 2-d double-well (\ref{eq:gradient}) in the space spanned by the first three diffusion coordinates, for parameter value $e=0.5$ far from the bifurcation (left) and for $e=1.5$ (right), which is closer to the bifurcation at $e\approx 1.73$. 
The lower panels {\bf b,c,e,f} show the same data in two-dimensional projections onto the diffusion coordinates ($\xi_1$,$\xi_2$) and ($\xi_1$,$\xi_3$). 
We use a lower noise level of $\sigma=0.09$ in order to obtain simulation data restricted to the shallow well when very close to the bifurcation, and thus the relation of $\xi_1$ and $\xi_2$ is different compared to Fig.~\ref{fig:grad2D_eigen_reconstr}e-f, where $\sigma=0.3$. 
}
\end{figure}

The leading $\xi_n$ are not necessarily all independent d.o.f's. In the system restricted to the shallow well, the critical d.o.f becomes the slowest upon approaching the bifurcation and the time scale separation with respect to all other d.o.f's becomes larger, at which point the first few backward eigenfunctions (and hence the first few diffusion coordinates) are all expected to parameterize the slowest d.o.f. This is particular for single-well systems. For instance, in a multi-dimensional parabolic potential with a slow variable $x$ and a spectral gap, $\xi_1$ is a function of $x$ and the next $k$ eigenfunctions $\xi_k$ (with $k$ dependent on the magnitude of the spectral gap) are polynomially related to $\xi_1$ \cite{NAD06,COI08}. %or more generally to xi1 and xi_n with n<k?
In this case, $\xi_1$ is sufficient as a reduction coordinate and is the only diffusion coordinate that indicates monotonically how far fluctuations evolve towards the saddle. 
Accordingly, in the 2-d double-well restricted to the shallow well, the dynamics in the space of the diffusion coordinates $\xi_{1,2,3}$ evolves from a two-dimensional to a one-dimensional manifold as the bifurcation is approached (Fig.~\ref{fig:grad2D_xi}). 

%Further away from the bifurcation, $\xi_2$ represents the slow mode in the $y$-direction, and the two other diffusion coordinates are quadratically related and parameterize the d.o.f related to the asymmetry of the potential well towards the edge state (Fig.~\ref{fig:grad2D_xi}a-c). 
%Close to the bifurcation, the dynamics becomes constrained to an approximately one-dimensional curve where $\xi_2$ ($\xi_3$) is a quadratic (cubic) function of $\xi_1$ (Fig.~\ref{fig:grad2D_xi}d-f). 
%The observed pattern formerly associated with $\xi_2$ at $e=0.5$ drops to higher eigenfunctions. 
%%% NOTE: it is actually at xi4!

\begin{figure}%[floatfix]%!htb
\includegraphics[width=0.99\textwidth]{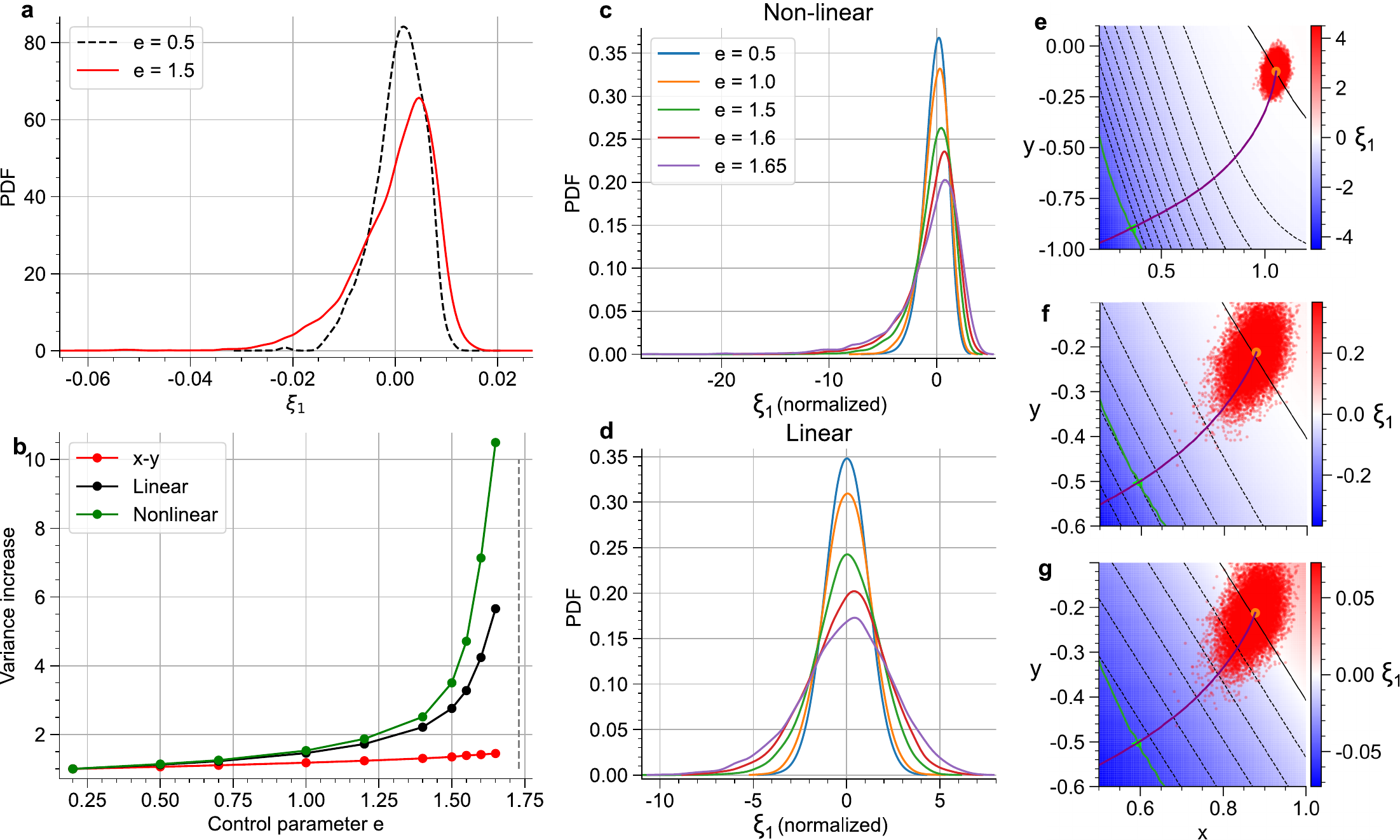}
\caption{\label{fig:grad2D_ews} 
Observables for EWS in the 2-d double well model (\ref{eq:gradient}). 
{\bf a} Distribution of the $\xi_1$-values evaluated at simulated data points using $e=1.5$ and $\sigma=0.09$ (red line). The black dashed line is a distribution of the same $\xi_1$ (i.e. obtained from the DM at $e=1.5$), but evaluated on data points simulated at $e=0.5$ via the Nyström interpolation (\ref{eq:nystrom})-(\ref{eq:nystrom2}). 
{\bf b-d} The diffusion coordinate $\xi_1$, estimated for simulation data close to the tipping point ($e=1.65$), is fit to a polynomial of the state variables (x,y), and used
as observable to detect CSD by evaluating it on residual data of simulations at lower values of the control parameter e. All simulations are initialized in the shallow well. Shown are distributions of the values of a linear ({\bf d}) and cubic ({\bf c}) polynomial, normalized to the fluctuations at $e=0.2$. 
Panel {\bf b} shows the variance increase of the linear and non-linear observable, as well as the observable $x-y$, normalized to the variance of the fluctuations at $e=0.25$. 
{\bf e-g} Polynomial fits of $\xi_1$ estimated from simulations (dots) with $\sigma=0.09$. Panel {\bf e} is a cubic fit to data at the parameter value $e=0.5$ far from the bifurcation, and ({\bf f,g}) is a cubic and linear fit for $e=1.5$, which is closer to the bifurcation. 
}
\end{figure}

%%% Extraction of observables as EWS. 
The observations above suggest three main ways to leverage information contained in the 
diffusion coordinates $\xi_n$ for early-warning of TPs. First, one may observe the qualitative change of the functional dependencies of the first $\xi_n$ as a result of the emerging time scale separation, as just discussed. 
Second, one can directly evaluate and compare $\xi_1$ for data sets obtained at different observational time slices (i.e. for different values of the control parameter) via the Nyström interpolation (\ref{eq:nystrom})-(\ref{eq:nystrom2}). In particular, we can estimate $\xi_1$ from a data set believed to be closest to a TP, for example from climate observations closest to present-day, and then evaluate the observable $\xi_1$ on data sets sampled further away from the TP, for example using climate observations of the past. If the variability and correlation of the values of $\xi_1$ is increased significantly in the former data set, this is an indication of decrease in resilience in the critical d.o.f and an impending TP. This is illustrated in Fig.~\ref{fig:grad2D_ews}a, where a clear change in variability is seen when evaluating $\xi_1$ estimated at $e=1.5$ on data simulated farther away from the bifurcation at $e=0.5$. Note that the same normalization of the variables that is applied before the DM algorithm to the data set where $\xi_1$ is estimated has to be applied to data at parameter values further away from the bifurcation. Here and in most of the following we focus our presentation on the variance as EWS, but similar plots could be shown for the autocorrelation. 

Third, an explicit expression of a physical observable can be constructed from $\xi_1$ as function of (possibly a subset of) the state variables. $\xi_1$ is estimated for an observational time slice closest to the TP and the values of $\xi_1$ at the data points (i.e. the entries of the eigenvectors of the Markov matrix $M$) are fit to a suitable function, e.g., a polynomial. The fitted function can then be evaluated for any data sets further away from the TP. 
In Fig.~\ref{fig:grad2D_ews}e-g we show polynomial fits to $\xi_1$ estimated from simulation data of the 2-d double-well. The directionality of the level sets is consistent with the direction of the edge state, as expected. Far from the bifurcation (assuming low noise) the dynamics samples only the relatively flat part of $\xi_1$ far from the saddle point, provided the noise is sufficiently small, as shown in Fig.~\ref{fig:grad2D_ews}e. 
Still, the fitted function shows a more rapid decrease towards the saddle point, thus indicating that it already carries the crucial information for detecting CSD. 
%maybe trivial?
When close to the TP, non-linear functions tend to be required for an adequate fit 
of the $\xi_1$ data (Fig.~\ref{fig:grad2D_ews}f). However, linear fits preserve the directionality of the edge state well (Fig.~\ref{fig:grad2D_ews}g). 
Using the fits estimated from data sampled close to the TP (Fig.~\ref{fig:grad2D_ews}f,g), we again see that the variability of the values of $\xi_1$ decreases when evaluated for data sample further away from the TP (Fig.~\ref{fig:grad2D_ews}c,d). The non-linear observable shows a significantly stronger change in variance compared to the linear one (Fig.~\ref{fig:grad2D_ews}b).

\begin{figure}%[floatfix]%!htb
\includegraphics[width=0.99\textwidth]{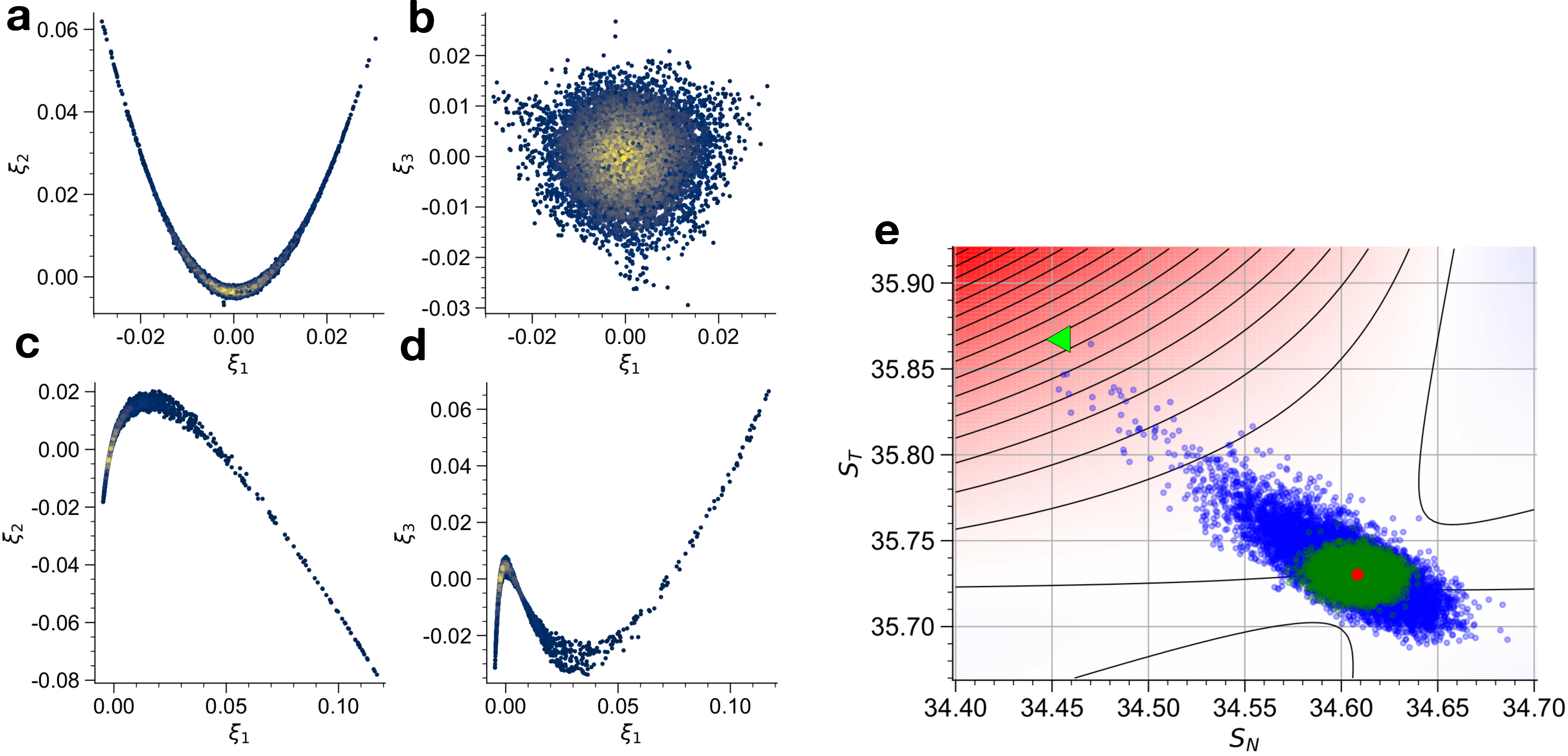}
\caption{\label{fig:fivebox_xi} 
{\bf a-d} Scatterplots of simulated data from the five-box AMOC model in projections onto the first three diffusion coordinates, for two values of the control parameter $H=0.25$ ({\bf a,b}) and $H=0.315$ ({\bf c,d}). 
{\bf e} Physical observable associated with the leading diffusion coordinate $\xi_1$, using as cubic polynomial of the variables $S_N$ and $S_T$, at $H=0.315$. 
The position of the edge state is marked by the green triangle, and the base attractor is the red dot. 
The blue point cloud is the data at $H=0.315$ used for the construction of the DM. The green point cloud is corresponding simulation data further from the bifurcation ($H=0.25$), shifted such that the mean is at the base attractor at $H=0.315$. 
}
\end{figure}

Similar results are obtained when applying the method to a slightly more complex non-gradient system, a four-dimensional conceptual model of the AMOC \cite{WOO19}, which is summarized in the following. The variables ($\{ S_N, S_T, S_S, S_I\}$) are the average salinities in four boxes of the global ocean (see Sec.~\ref{AppB} for the equations). 
Large differences in the box volumes gives a time scale separation, with $S_S$ and $S_I$ being the slowest variables, and $S_N$ being the fastest. The model is bistable for a range of the control parameter $H$ from $H\approx 0.04$ until the bifurcation at $H\approx 0.3214$, where the stable fixed point corresponding to a present-day AMOC disappears. 

We use simulation data restricted to the present-day AMOC state. Far from the bifurcation, $\xi_1$ represents a correlated relaxation mode in the slow variables $S_S$ and $S_I$. $\xi_2$ is a quadratic function of $\xi_1$, and $\xi_3$ is independent and strongly correlated with $S_N$ only (Fig.~\ref{fig:fivebox_xi}a,b). When approaching the TP, the slow mode corresponding to relaxation along the direction of the edge state
emerges due to CSD, and eventually rises to the position of $\xi_1$, 
with $\xi_2$ ($\xi_3$) being a quadratic (cubic) function thereof (Fig.~\ref{fig:fivebox_xi}c,d). $\xi_1$ exhibits a strong non-linear anti-correlation with $S_N$ and a non-linear positive correlation with $S_T$ (see also \cite{LOH25}). A good physical observable representing $\xi_1$ is found by considering polynomial functions in a projected space of a subset of (e.g. two) variables. The best cubic polynomial fit of $\xi_1$ (estimated at $H=0.315$) is a function of $S_N$ and $S_T$. The resulting function shows a monotonic, non-linear increase from the base state towards the edge state, while being essentially flat in other directions (Fig.~\ref{fig:fivebox_xi}e). This renders the observable very sensitive to excursions towards the edge state, and thus ideal for EWS.

\subsection{Observables and extrapolation of tipping times}
\label{sec:extrapolation}

So far we discussed the construction of observables that show a large increase in variance (and also autocorrelation) as a result of CSD, which is a qualitative indication that the system moves towards a TP. One may go further and attempt a quantitative prediction of the expected time of tipping by extrapolating the CSD signal in observational time series, as was done in the context of real-world climate observations in \cite{BOE21,DIT23}. For this one needs to assume that an observed time series samples the critical dynamics by obeying the SNB normal form. This implies that such a prediction is sensitive to the choice of observable, as shown in the following. 

Consider the general multi-dimensional system described by the coupled stochastic differential equations (\ref{eq:system}), where we now assume that the drift $b^{\gamma}(\mathbf{X}, \mu)$ depends on a control parameter $\mu$ and the noise $\sigma_{\nu}^{\gamma}(\mathbf{X}) = \sigma_{\nu}^{\gamma}$ is additive. If the system undergoes a SNB, the noise-driven dynamics is expected to become restricted to the vicinity of a one-dimensional center manifold, and is described by the normal form 
\begin{equation}
dx = (x^2 - \mu)dt + \sigma dW_t, 
\end{equation}
where $\mu = 0$ demarcates the bifurcation. 
Close to the fixed point, the system can be linearized and approximated by the Ornstein-Uhlenbeck process $d\tilde{x}_t = -\lambda \tilde{x} dt + \sigma dW_t$. Crucially, the linear restoring rate $\lambda$ is related to the control parameter with $\lambda = 2 \sqrt{\mu}$. 
%%% lambda related to real part of largest Jacobian eigenvalue!
Data sampled at small time intervals $\Delta_t$ can be approximated by an AR(1) process 
\begin{equation}
X_{k+1} = e^{-\lambda \Delta t} X_k + \epsilon_k,
\end{equation}
where $\epsilon_k$ are Gaussian random variables with variance $\sigma^2 (2\lambda)^{-1} (1-e^{-2\lambda \Delta t})$. For this process the autocorrelation at lag 1 is given by $\rho_1 = e^{- \lambda \Delta t}$. Since $\lambda = 2 \sqrt{\mu}$, we can reconstruct $\mu$ from data by
\begin{equation}
\label{eq:ctrl_param}
\mu = \left( \frac{\ln \rho_1}{2 \Delta t} \right)^2. 
\end{equation}
Thus, in a sliding window one can estimate $\rho_1$ as a function of time, and, assuming a linear trend in $\mu$, estimate with a linear fit to the function on the righthand-side at what time the control parameter will cross zero, which is exactly when the autocorrelation tends to 1 at the SNB.

\begin{figure}%[floatfix]%!htb
\includegraphics[width=0.5\textwidth]{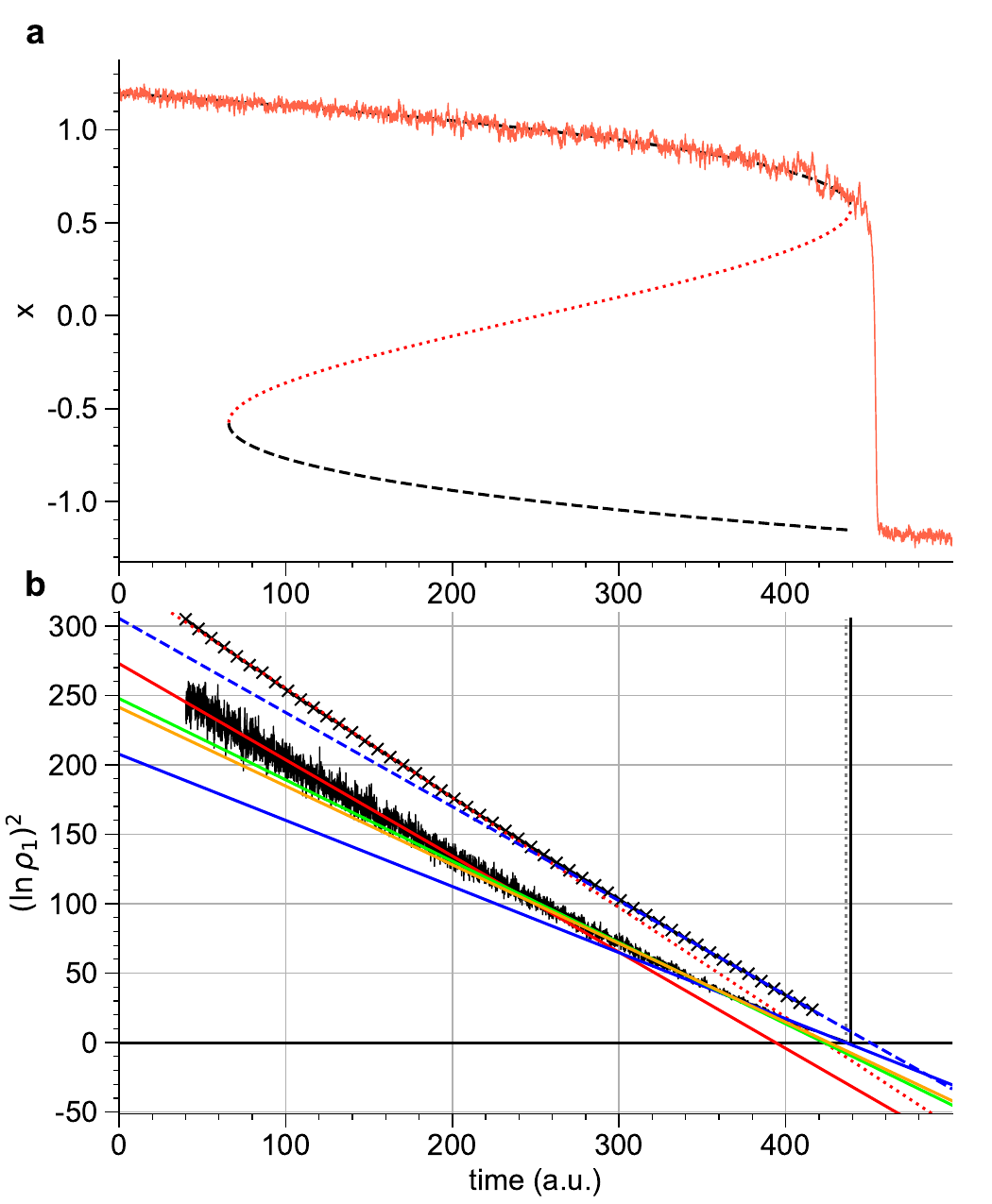}
\caption{\label{fig:tipping_time_dw1} 
Simulations of the 1-d double well (\ref{eq:dw1}) with $\sigma = 0.05$ and a linear increase of the control parameter within 500 time units from $\beta = -0.7$ past the bifurcation point $\beta_c = \sqrt{8/27}$ to $\beta = 0.7$.  
{\bf a} Time series of one simulation overlaid on the bifurcation diagram.
{\bf b} Reconstructed control parameter (\ref{eq:ctrl_param}) obtained by the lag-1 autocorrelation $\rho_1$, estimated at each time step ($\Delta t = 0.05$ time units) for an ensemble of 15,000 simulations (black trajectory) until a cutoff time where 98\% of the ensemble members have not tipped yet (evaluated by crossing a threshold of $x=0.2$). At the noise level $\sigma = 0.05$, this is only very shortly before the bifurcation is reached. The 2\% of realizations that tipped are removed. The solid lines are linear fits using different segments of the data, with the red (blue) line using the first half (last sixth) of the data. The crosses as well as dotted and dashed lines are for $\rho_1$ estimated in a moving window of length 40. 
}
\end{figure}

If given data from an arbitrary scalar observable of the system, this extrapolation to the time of tipping can fail since the observable need not obey the SNB normal form, or may only approximately do so when arbitrarily close to the bifurcation. 
In fact, even for univariate bi-stable systems (where the question of observable is obsolete), the scaling $\mu \propto [\ln (\rho_1(X))]^2$ according to the saddle-node normal form only applies when close to the bifurcation. 
For the 1-d double-well (\ref{eq:dw1}) under a linear change in time of the control parameter $\mu$ the function $[\ln (\rho_1(X))]^2$ is convex with respect to time (Fig.~\ref{fig:tipping_time_dw1}). 
%, i.e., derivative increasing towards 0 from negative values, 
Thus, the TP, which occurs at $t\approx 420$, would be predicted too early at $t\approx 390$ (red line in Fig.~\ref{fig:tipping_time_dw1}b) when the linear extrapolation is performed based on data not close enough to the bifurcation. An estimation of the autocorrelation in a moving window introduces a further bias towards a later estimated tipping, because the autocorrelation is underestimated by removing some correlation during the necessary step of detrending within each window. 

\begin{figure}%[floatfix]%!htb
\includegraphics[width=0.99\textwidth]{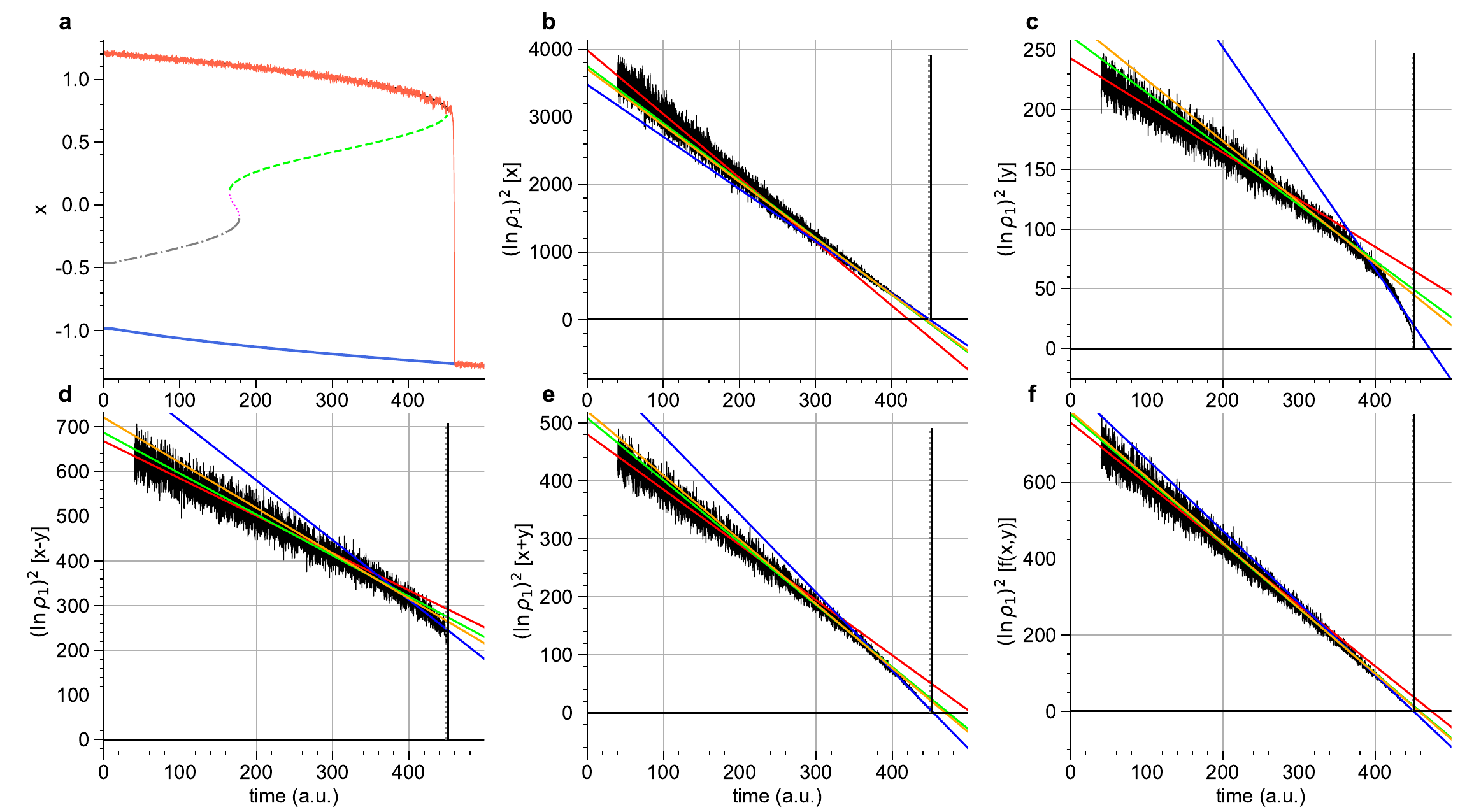}
\caption{\label{fig:tipping_time_dw2} 
Same as Fig.~\ref{fig:tipping_time_dw1}, but for simulations with the 2-d double-well (\ref{eq:gradient}) with $\sigma_x = \sigma_y = 0.05$, where the parameter $e$ is increased linearly past the bifurcation. Shown in {\bf a} is a single example trajectory overlaid on the bifurcation diagram. Panels {\bf b}-{\bf f} show the evolution of the quantity $[\ln (\rho_1(O))]^2 (2\Delta t)^{-2}$ for different observables $O(x,y)$, along with linear fits to different parts of the time series. 
%NOTE: perhaps need more detail, as in the figure caption before. 
}
\end{figure}

As more dimensions are involved the prediction depends on the choice of observable. For the 2-d double-well (\ref{eq:gradient}), the variable $x$ can give an accurate prediction when data are available close enough to the TP (Fig.~\ref{fig:tipping_time_dw2}b). 
In contrast, $y$ initially shows a quasi-linear relation of $\mu$ and $[\ln (\rho_1(X))]^2$, but then a much steeper relationship closer to the bifurcation (Fig.~\ref{fig:tipping_time_dw2}c). 
Extrapolating of the initial slope would lead to a tipping time estimate that is far too late. Especially unsuited observables exist, such as $O(x,y) = x-y$, where no tipping can be predicted before a noise-induced transitions would occur (Fig.~\ref{fig:tipping_time_dw2}d). 
In contrast, $O(x,y) = x+y$ is very closely aligned with the direction of the edge state \cite{LOH25} and permits to predict the time of tipping accurately for data sufficiently close to the bifurcation (Fig.~\ref{fig:tipping_time_dw2}e). The non-linear observable obtained from the DM approximation to the first subdominant backward eigenfunction (Fig.~\ref{fig:grad2D_ews}f) is most accurate, even when evaluated at data far from the bifurcation (Fig.~\ref{fig:tipping_time_dw2}f).

%%% different aspects: ``wrong'' observable can still work. possibly because different effects cancel out. 

%%% Ideas on why it can fail:

%%% It is based on linearization. -> not sure nonlinear observables are useful here. 
%%% possibly assumes we are in tangent space of central manifold. 
%%% and on top of that maybe only valid for very LOW NOISE. 
%%% NOTE: there is the possibility that all of the observables have a bias to predicting 
%%% too early, due to finite noise. 
%%% NOTE: kind of makes sense, because at least in 1D picture the well is actually clearly
%%% less steep in saddle direction than what parabolic fit would suggest!!!
%%% NOTE: then it should also be a bias in the 1D model...

%%% NOTE: using observables that are projected down we also cannot claim they are 
%%% correct at predicting tipping time!

\subsection{Application to tipping points in a global ocean model}
\label{sec:veros}

\begin{figure}%[floatfix]%!htb
\includegraphics[width=0.5\textwidth]{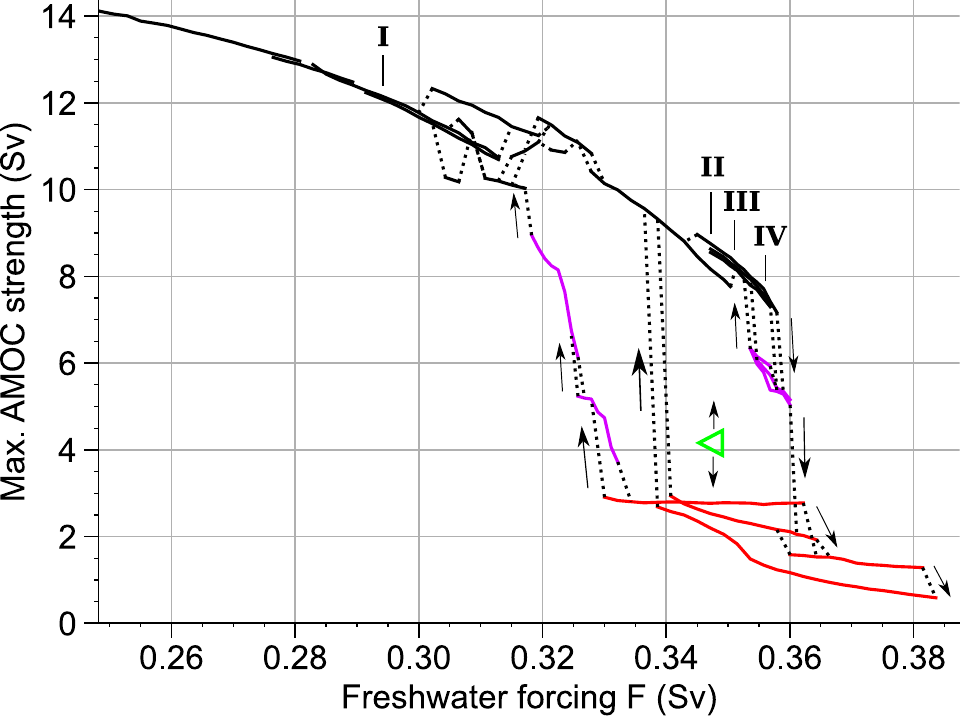}
\caption{\label{fig:veros_bif} 
Bifurcation diagram of the Veros ocean model without noise forcing (obtained in \cite{LOH24}), with the maximum AMOC strength as order parameter, and the freshwater forcing $F$ as control parameter. All individual solid lines correspond to different branches of attractors, and the dotted lines as well as the arrows indicate the transition path of the system as a given attractor loses stability. The edge state at $F=0.3472$ (computed in \cite{LOH24b}) is marked by the green triangle. 
}
\end{figure}

As the final result, we show our method is capable of successfully detecting TPs 
in a high-dimensional system. We consider the global ocean model Veros \cite{HAE18}, which shows a TP of the AMOC from its present-day state to a collapsed state as a result of increasing meltwater input to the North Atlantic. Veros is a primitive-equation finite-difference ocean model forced with a fixed atmospheric climatology, and discretized on a grid of 40 latitudinal and 90 longitudinal grid points, as well as 40 depth levels. This is a coarse-resolution setup, but it enables long steady-state simulations to get good statistics beneficial for our feasibility study. As a dynamical system, the model possesses almost one million degrees of freedom. For more details on the model, see \cite{HAE18,LOH21,LOH24}. 
The meltwater input $F$ is the control parameter, and the stability landscape with respect to $F$ (computed in \cite{LOH24}) is shown in Fig.~\ref{fig:veros_bif}. There are several branches of attractors with an AMOC similar to present-day, but these collapse at 
a high freshwater forcing of $F \approx 0.36$. After this TP, there remain only attractors with a collapsed AMOC.

\begin{figure}%[floatfix]%!htb
\includegraphics[width=0.9\textwidth]{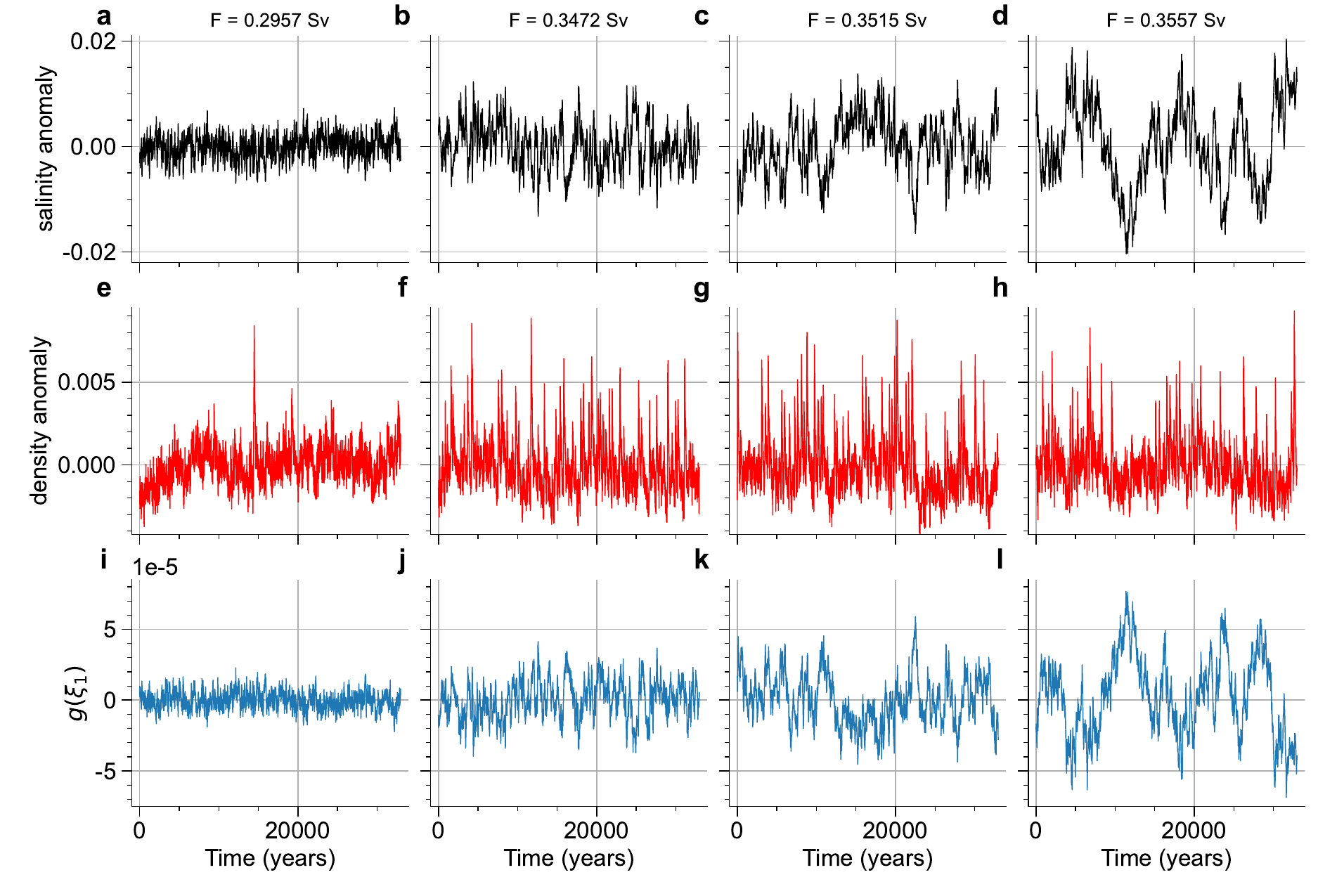}
\caption{\label{fig:veros_timeseries} 
{\bf a-h} Example time series of two variables in the four Veros data sets used here. Shown is the north Atlantic deep ocean salinity ({\bf a-d}) at the four different values of the control parameter $F$, as well as the south Atlantic deep ocean density ({\bf e-h}). The time series are shown as anomalies with the mean removed. 
{\bf i-l} Time series of an observable constructed by projection onto the leading diffusion coordinate $\xi_1$ in the subspace of deep ocean salinity. The eigenmode was estimated with the DM algorithm from data sampled at $F=0.3557$~Sv. %NOTE: more details on the projection?
}
\end{figure}

We use four 33,000-year long equilibrated simulations performed at four, fixed values of $F$ leading up to the TP, and sampled as 5-year averages of the state variables. These are referred to as simulations I to IV, see Fig.~\ref{fig:veros_bif}.
Additive surface temperature and salinity noise drives fluctuations of the system around its deterministic attractors (for more details see \cite{LOH25}), which otherwise feature relatively small-amplitude chaotic oscillations \cite{LOH24}. 
The system without noise forcing has been investigated previously to determine an edge state on the separatrix of the present-day and collapsed AMOC regimes \cite{LOH24b}. By analyzing its mean climatology it was found that the edge state distinguishes itself most strongly from the mean states on the attractors in terms of its fresh and cold deep Atlantic. This is the ``fingerprint'' of the edge state. Subsequently, it was argued that increased fluctuations towards the edge state as a result of CSD should be most prominent in variables quantifying this fingerprint \cite{LOH25}. 
%i.e., where it is moving most as the control parameter is changed.
%%%NOTE: REFINE THIS THOUGHT
%%% maybe point out here that sensitivity to control parameter would presumably
%%% mean large fluctuations?
%edge state and base attractor are most sensitive to the control parameter change. 
%strongest change in the quasipotential in this direction in phase space. 
Indeed, only a very small subset of all d.o.f, coinciding exactly with the variables describing the deep ocean fingerprint, shows significant variance increase prior to the AMOC collapse \cite{LOH25}. The variable that was found to exhibit the largest increase in variance is shown in Fig.~\ref{fig:veros_timeseries}a-d across the four data sets. 
In the following, we show that similar (if not better) results can be obtained with our 
the DM method which does not require prior knowledge of the edge state or a brute-force search across all d.o.f (risking false positives). Instead, only observational data close to the TP is required. 

It would be feasible to compute the DM distance Kernel in the full space of the three-dimensional fields, perhaps after a weighting of the different physical units (temperature, salinity, density and velocity). But for simplicity we perform an initial dimensionality reduction, by averaging the salinity, temperature and density fields over boxes covering the entire ocean at different depths, and by summarizing the strength of the main ocean currents in terms of the spatial maxima of the meridional and barotropic stream functions (see \cite{LOH24b}). This yields time series of 83 variables covering most important aspects of the model state. Fig.~\ref{fig:veros_timeseries}a-h shows time series of two of these variables at four different values of $F$, with the mean removed. Before applying the DM algorithm, we normalize all variables to have unit variance. 
A bandwidth of $\epsilon = 17$ for the Kernel (\ref{eq:kernel}) was found to be optimal to resolve the data manifold at all parameter values $F$ without being influenced by outliers. 
%large epsilon will smear out the data and destroy the structure of the data manifold. 

\renewcommand{\arraystretch}{0.8}
\begin{table*}[!htb]%[width=0.5\textwidth]
 \setlength{\tabcolsep}{4pt}
\centering
\caption{\label{tab:diffmap_results} Spearman correlation of the top five Veros variables with the first two diffusion coordinates $\xi_1$ and $\xi_2$, where the $\xi_n$ were estimated at the four different values of the control parameter $F$. A shorthand notation encodes the meaning of the variables. E.g., 'rho deep TP' is deep ocean density in the tropical Pacific. The first part ('rho', 'salt' or 'temp') refers to density, salinity or temperature, respectively. The second part ('subs' or 'deep') refers to the subsurface ocean averaged until 1000m depth or the deep ocean below 1000m.  
The last part is geographical location, which can be the northern, tropical or southern Atlantic or Pacific (NA, TA, SA, NP, TP, SP), or the Southern Ocean (SO) and the Indian Ocean (IO). In bold text are key variables related to the mode of cold and fresh excursions in the deep north and tropical Atlantic directed towards the edge state. In italic are key variables related to the mode of fast cold excursions in the Southern ocean, which strongly increase the density in large parts of the deep ocean.}
%\resizebox{\textwidth}{!}{%
{%
\begin{tabular}{llr|lr|lr|lr}%
 \hline
 & \textbf{F=0.2957} & & \textbf{F=0.3472} &  & \textbf{F=0.3515} &  & \textbf{F=0.3557} &  \\
  & Variable & \textbf{$r_S$} & Variable & \textbf{$r_S$} & Variable & \textbf{$r_S$} & Variable & \textbf{$r_S$}  \\
 %\midrule
  \hline
$\xi_1$	&rho deep TP& 0.854&{\it rho deep SA}&	-0.874	&{\it rho deep SA}&	-0.886&	{\bf salt deep NA}&	-0.873\\
&	rho deep TA&	0.848&	rho deep TA&	-0.870	&rho deep TA&	-0.868&	{\bf salt subs NA}&	-0.834\\
&	{\it rho deep SP}&	0.835&	{\it rho deep IO}&	-0.842	&{\it rho deep IO}&	-0.860&	{\bf temp deep NA}&	-0.789\\
&	{\it rho deep IO}&	0.822&	{\it rho deep SP}&	-0.803	&{\it rho deep SP}&	-0.827&	{\bf salt deep TA}&	-0.735\\
&	{\it rho deep SA}&	0.813&	{\it rho deep SO}&	-0.794	&{\it rho deep SO}&	-0.821&	temp deep IO&	0.730\\
  \hline
$\xi_2$	 &temp subs TP&	 0.627&	temp subs TA&	0.624&	{\bf salt deep NA}&	-0.775&	{\it rho deep SO}&	-0.782\\
&	rho subs TP&	-0.607&	rho subs TA&	-0.615&	{\bf salt subs NA}&	-0.765&	{\it rho deep SA}&	-0.763\\
&	rho subs TA&	-0.597&	rho subs TP&	-0.567&	{\bf temp deep NA}&	-0.690&	{\it rho deep SP}&	-0.723\\
&	temp subs TA&	0.582&	    salt subs NA&	-0.563&	{\bf temp subs TA}&	0.667&	  salt deep SO&	-0.701\\
&	temp deep TP&	0.518&	    temp subs TP&	0.554&	    rho subs TA&	-0.643&	rho deep TP&	-0.681\\
 \hline
\end{tabular}}
\end{table*}

The physical meaning of the first two inferred diffusion coordinates is summarized in Tab.~\ref{tab:diffmap_results}, where the five physical variables with the highest correlation to $\xi_1$ and $\xi_2$ are listed. As the control parameter is changed towards the TP, we can see that the expected critical mode emerges. When far from the TP, $\xi_1$ is best correlated with deep ocean density in the Indo-Pacific, South Atlantic and Southern Ocean. The next mode, represented by $\xi_2$, is best explained by variability in the tropical subsurface ocean. When increasing the control parameter to $F=0.3515$, it is replaced by a mode correlated with temperature and salinity in the deep northern and tropical Atlantic. Increasing $F$ further to $F=0.3557$ shortly before the TP, this becomes the leading mode $\xi_1$. The most important variables in this mode (upper right column in Tab.~\ref{tab:diffmap_results}) are exactly those that make up the fingerprint of the edge state \cite{LOH24b} and feature the largest increase in variance \cite{LOH25}. 

\begin{figure}%[floatfix]%!htb
\includegraphics[width=0.48\textwidth]{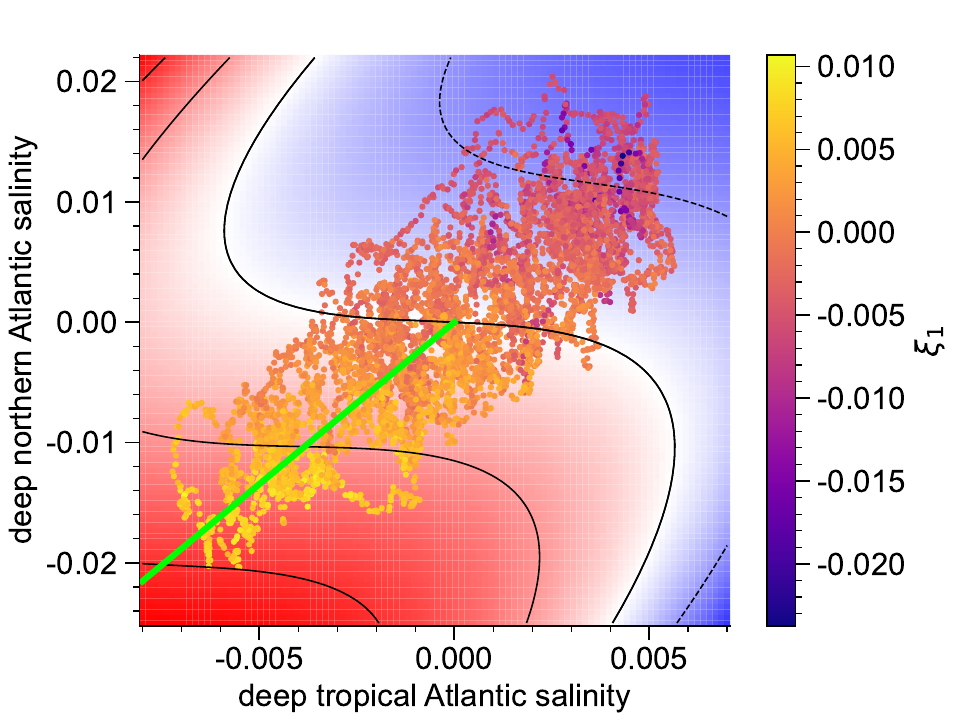}
\caption{\label{fig:veros_observable_2d} 
Two-dimensional observable (color map and contours) created from a cubic polynomial fit of two variables of the Veros data (north and tropical Atlantic deep ocean salinity at parameter value $F=0.3557$) to the values of $\xi_1$ obtained from the diffusion map. The point cloud depicts the data with color coding according to corresponding the value of the diffusion coordinate $\xi_1$. Shown are anomalies with respect to the mean state of the model. The green line is a vector pointing from this mean state to the edge state \cite{LOH25}. 
}
\end{figure}

To design a physical observables for EWS from $\xi_1$, the first option is again to use the Nyström extension based on all variables and interpolate the function $\xi_1$ to observations further back in time, in order to find evidence for increased fluctuations in the critical mode. 
One may also fit $\xi_1$ as a linear or non-linear function of the variables. In high dimensions it is sensible to only consider a subset of variables to find a parsimonious observable with the best signal-to-noise ratio when applied for EWS. 
We leave a treatment of this statistical optimization problem for future research, and consider here three simple examples. First, simply take the variable with highest correlation to $\xi_1$ as observable. This is north Atlantic deep ocean salinity (Fig.~\ref{fig:veros_timeseries}a-d), which indeed has the highest increase in variance of all individual features \cite{LOH25}, increasing by a factor of 12.23 when going from $F=0.2957$ to $F=0.3557$. %actually the first 2 features have highest increase in variance!
Second, fit $\xi_1$ to a linear combination of the two best features in Tab.~\ref{tab:diffmap_results} (at $F=0.3557$). This yields a variance increase by a factor of 14.21. 
Third, fit $\xi_1$ to a cubic polynomial of two variables, which are chosen as the two features among the best four (Tab.~\ref{tab:diffmap_results}) that represent two different ocean sectors, thus giving some degree of spatial independence. This observable captures the directionality of the edge state well (Fig.~\ref{fig:veros_observable_2d}), and shows an increase in variance by a factor of 12.60. 

\begin{figure}%[floatfix]%!htb
\includegraphics[width=0.99\textwidth]{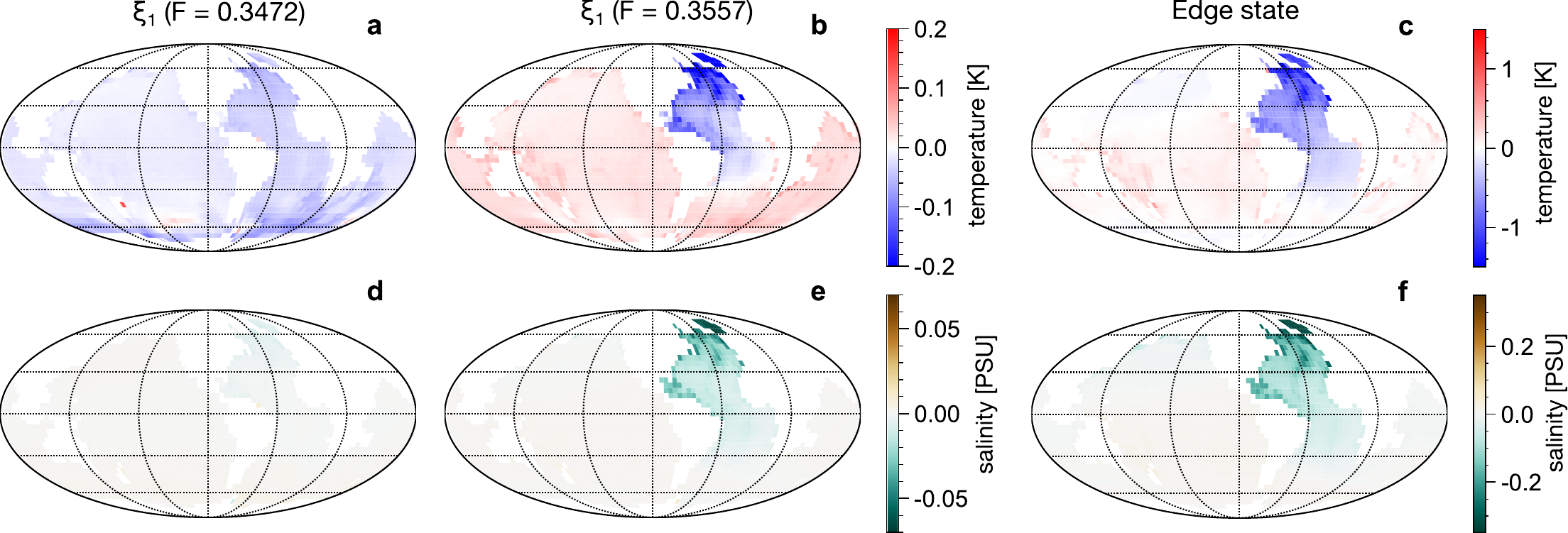}
\caption{\label{fig:veros_modes} 
{\bf a,b,d,e} Anomaly maps of deep ocean (vertical average below 1000m) temperature ({\bf a,b}) and salinity ({\bf d,e}) describing the first subdominant spatiotemporal mode $\xi_1$ of the Veros model obtained by the DM algorithm at the control parameter values $F=0.3472$ ({\bf a,d}) and $F=0.3557$ ({\bf b,e}). 
{\bf c,f} Corresponding anomaly map of the edge state with respect to the mean state on an attractor with active AMOC (state II in Fig.~\ref{fig:veros_bif}), as estimated from the deterministic version of the model in \cite{LOH24b}. 
}
\end{figure}

Finally, $\xi_1$ can be mapped back into the physical space of full dimension, whereafter spatio-temporal anomalies from different time periods can be projected onto the mode. 
In particular, we propose to obtain the physical representation of this mode by averaging over the time points where $\xi_1$ is extremized. For instance, one can choose the data points with the top 5\% largest and 5\% smallest values of $\xi_1$. 
Averaging independently over these two sets of data points defines a positive and a negative phase of the mode. By taking the difference of the positive and negative patterns we obtain a pattern that describes the mode as a whole and that we can project data onto. This may be viewed as linear approximation that interpolates the physical mode linearly as a function of the value of $\xi_1$. 
Fig.~\ref{fig:veros_modes} shows the first subdominant mode extracted in this way, comparing results far from (Fig.~\ref{fig:veros_modes}a,d) and close to the TP (Fig.~\ref{fig:veros_modes}b,e). The modes are projected down to the two-dimensional physical space of vertically averaged deep ocean of temperature and salinity. 
Far from the TP, the mode is characterized by a global cooling of the deep ocean initiated by abrupt (decadal-scale) cooling events in the Southern Ocean (seen as spikes of density increase in Fig.~\ref{fig:veros_timeseries}e-h), which are excited by noise in the multistability regime, as discussed in \cite{LOH25}. The resulting dense deep ocean water spreads throughout the deep ocean, before the signal decays on a multi-centennial time scale. There is no significant salinity signal. 

Close to the TP, the new physical mode that has emerged as subdominant eigenfunction is characterized by a cold and fresh anomaly of the deep northern and tropical Atlantic (Fig.~\ref{fig:veros_modes}b,e), with a spatial pattern that very closely resembles the anomaly of the edge state (Fig.~\ref{fig:veros_modes}c,f). 
A scalar observable is created by projecting the snapshots of the data fields (at each time step, and as anomalies with respect to a mean state) onto this pattern via the scalar product of the two fields. Comparing different observational slices (such as the simulations I to IV) then reveals the changes in variability of this critical mode, which serves as EWS. The variance increases by a factor of 21.05 when going from state I to state IV, and the time series of this observable are shown in Fig.~\ref{fig:veros_timeseries}i-l. 

While these are encouraging results, in practice there is a risk of false positives and negatives, since here the critical physical mode only appears in its correct place at $\xi_1$ when already quite close to the TP. One needs to verify that the leading mode captured by $\xi_1$ is likely a critical mode. The guiding principle should be to look how strongly fluctuations in the leading diffusion coordinate, estimated at the current time slice, have increased compared to data slices back in time, and then set a level of statistical significance based on a reference period. 
In the Veros data one can in this way rule out the initially leading mode (Fig.~\ref{fig:veros_modes}a,d), because its variability does not keep increasing towards the TP (Fig.~\ref{fig:veros_timeseries}f-h), and the associated excursions do not last longer. But in principle there remains a chance for false positives if a new time scale separation would arise upon change of a control parameter for reasons not related to a TP. 

%NOTE: could also make a statistical procedure, where one estimates which variables 
% in the extremized xi1 average is significantly different from the variability 
% on the attractor. 

%NOTE: could be worthwhile in a Supp figure to show that indeed xi2 and xi1 switch,
% by showing correspondence of the fields. 
%NOTE: could also add Supp figure where we show that surface+subsurface xi1 pattern 
% ALSO a signal, but just NA subsurface cooling and TA warming + NA freshening. 
% so it is the same phenomenon. 

%NOTE: other idea for maybe better fits and statistics:
%%% first compute the marginal percentiles of xi1 and then fit to observables. 

%%% variance increases: 
%%% model with best 2 features: 14.21
%%% linear model salt_deep_TA and salt_deep_NA: 13.44
%%% nonlinear model of the same: 12.6
%%% projection onto extremized xi1 in deep ocean salintiy: 21.05

\section{Discussion}
\label{sec:discussion}

%%% short SUMMARY of method/approach
Here we propose a method to obtain observables for detecting CSD before bifurcation-induced TPs \cite{ASH12} in multi-stable systems via a data-driven approximation of the FP operator. 
The scenario of bifurcation-induced tipping implies low noise, where the first $k$ FP eigenfunctions $\psi_k$ are very slowly decaying modes related to rare noise-induced escape between metastable states \cite{HUI04}. The subsequent eigenfunctions describe probability density patterns in phase space that relax slowest towards a quasi-stationary distribution centered around a metastable state. We consider without loss of generality the bistable situation with $k=1$, where at the TP a base state loses stability and the system transitions to its alternative state.
In practical applications, the system prior to the TP is only observed in the basin of the base state and the contribution of $\psi_1$ remains quasi-constant (until infinitesimally close to the TP). 
In this case, the first non-trivial eigenfunction that can be observed in data is 
$\psi_2$, which describes the slowest relaxation mode towards the quasi-stationary distribution around the base state within its basin of attraction. As the TP is approached, due to CSD this mode will eventually represent the slowing relaxation along the critical d.o.f. 

We suggest to approximate the corresponding eigenfunction of the backward FP operator by the first diffusion coordinate $\xi_1$, obtained as scaled eigenvector of the DM Markov matrix. From $\xi_1$, one can obtain a physical observable - e.g. by projecting onto patterns obtained as average over data points that extremize $\xi_1$ - that shows a monotonic increase from the base state along the critical d.o.f towards the edge state. 
With several examples of low-dimensional bi-stable models we demonstrated that such an observable shows increases in noise-driven fluctuations that provide robust statistical EWS of the CSD associated with the impending TP. 
We also showed that measuring CSD in the correct observable is crucial when attempting to
predict the time of tipping by extrapolating the scaling of variance or autocorrelation of a scalar time series based on the SNB normal form. 
The method was successfully applied to simulation data from a high-dimensional global ocean model that features a TP of the AMOC. A critical mode was extracted that is in excellent agreement with the mode that would be expected from knowledge of the edge state \cite{LOH24b,LOH25}, and a scalar observable was derived from $\xi_1$ that shows highly significant increases in variance that are useful as statistical EWS. 

%%% CAVEATS: general for all methods. 
%% 1. need to be close to TP
A general caveat for this and other methods aiming to measure CSD from high-dimensional systems is that the critical mode may only emerge as the first subdominant eigenmode when already very close to the TP. 
If there are competing non-critical modes % NON-BIFURCATING?
that are very slow, it may not be possible to identify the critical mode, since one would need a very long observational time horizon while being close enough to the TP where the critical mode is finally the slowest. 
Indeed, it was seen for the Veros model that the correct physical mode takes on the role of $\xi_1$, but only as the system was quite close to the TP. Unless one knows beforehand which d.o.f should be measured, for instance by knowledge of the edge state \cite{LOH25} or robust physical considerations, it may only be possible to issue a reliable warning when very close to the TP, and potentially only after the probability of noise-induced transitions has become substantial. 

% Relation to other methods (regarding EWS). 
There are other previously proposed dimensionality reduction methods aiming to extract a scalar observable that can be used to detect CSD. These include variance-based techniques \cite{HEL04,BAT13,KWA18,PRE19}, where the first principle component is identified by empirical orthogonal functions - as originally proposed to obtain the critical mode for EWS \cite{HEL04} - or principal oscillation patterns. 
An autocorrelation-based method has also been proposed, where the directions of 
maximum variance of the first differences of multivariate time series are found, which  gives components of high autocorrelation that should indicate directions of lowest resilience \cite{WEI19}. 
% Methods that keep analyzing the full multivariate data set, and 
Other methods look for a SNB 
%(change in local stability around moving fixed point) 
in the full set of (observed) variables via the eigenvalues of a reconstructed Jacobian, which is determined by fitting a multivariate autoregressive model \cite{WIL15} or a multidimensional Langevin equation \cite{MOR24}. %NOTE ONLY SNB!
% definitely reasonable methods; but may become hard (data hungry) when working in 
% high-dim phase space. 
%MOR24 kind of nonlinear, but then uses linearization (Jacobian) 

Our approach is distinguished by combining several attributes. It yields an observable 
derived from the first subdominant backward FP eigenfunction that is designed to represent the critical mode displaying CSD, based on the flattening of the quasipotential expected for a broader class of TPs. This mirrors the reasoning for a natural tipping observable recently proposed in \cite{LUC24}, and it is also supported by other operator-theoretic work on the topic \cite{TAN18,CHE20,TAN20,GUT22,ZAG24}. 
The DM algorithm that we propose to use is a non-linear dimensionality reduction method and thus allows for observables to be non-linear functions of the state variables. 
It can be deployed for relatively high dimensional systems, since the quality of its approximation for a given sample size is not dependent on the full state space dimension, but on the intrinsic dimension of the underlying data manifold, which may be much lower. An equal time spacing of data points, or any time ordering at all, is not required, thus increasing its applicability for instance to paleoclimate data \cite{LOH25c}. 
The method furthermore allows for a qualitative assessment of changes in the dominant physical modes, by observing the functional relation between eigenfunctions, and the relation of eigenfunctions and physical variables. This is useful for determining an emerging time scale separation before the TP, and it may help to prevent false positives for instance due to changing amplitude or correlation of the driving noise. Estimating the eigenfunctions with the generalized DM method presented in \cite{BAN20} also allows treating multiplicative noise and non-gradient dynamics, which can alleviate the issues that unknown changes in oscillatory modes cause for statistical EWS \cite{LOH24}. This method gives equivalent results when applied to the ocean model (see App.~\ref{AppA}). 

%%% OUTSTANDING ISSUES / FUTURE WORK. 
Future work should address how systems driven far away from a steady state need to be treated. In many cases of real-world relevance, such as tipping of the polar ice sheets, the change in the control parameter is fast compared to the critical relaxation mode $\psi_2$ and perhaps many other modes. This means that the system is not in a quasi-stationary state, as was assumed here, and it is likely that the critical mode is not displayed before crossing the TP. An extended framework based on non-autonomous dynamical systems theory can hopefully yield useful insights on whether EWS before the de-facto TP exist in this case. %which may be delayed. 
%NOTE: also mention R-tipping as additional, genuine non-autonomous instability, and PNAS 2021 paper. 

% commitor functions, even though one would need to know direction of edge state at least 
%NOTE: slightly alternative approach would be to construct ``local'' committor function from diffusion maps 
% by adding boundary conditions to domain around base attractor and escape domain. 
% on one hand this is good, because committor is in certain way ideal as observable for the escape. 
% on the other hand, we do not a priori know where to put the escape domain, and it would be easy 
% to put it in the wrong place (for instance for the Veros data). 
% could work if one first identifies the metastable states from the eigenfunctions, with the saddle being 
% the alternative state, where eigenfunction is also constant. 

%NOTE: something about Koopman (modes), and e.g. dynamic mode decomposition?

%NOTE: another generalization: anisotropic diffusion map with weighted euclidian metric?
%      this is of course difficult for system where we don't know noise structure. 

%NOTE: Further alterations: use symmetric matrix adjoint to M.
%      Variable bandwidth kernel. 

%\bibliographystyle{apsrev4-2} % Tell bibtex which bibliography style to use
\bibliography{refs} % Tell bibtex which .bib file to use (this one is some example file in TexLive's file tree)

\begin{acknowledgments}
GAG acknowledges valuable discussions with Matheus Mankato de Castro and Caroline Wormell. We thank R. Nuterman and the Danish Center for Climate Computing for supporting the
simulations with the Veros ocean model. J.L. has received support from Danmarks Frie Forskningsfond (grant no. 2032-00346B) and research grant no. VIL59164 from VILLUM FONDEN.
\end{acknowledgments}

\appendix 
\setcounter{table}{0}
\renewcommand{\thetable}{A\arabic{table}}

\section{Eigenfunctions in non-gradient systems with non-additive noise estimated by local Kernel diffusion maps}
\label{AppA}

Here we present how the Fokker-Planck and backward Kolmogorov eigenfunctions can be estimated without assuming gradient dynamics or additive noise, and demonstrate that similar results are obtained for data from the Veros model as with the standard diffusion map method presented in the main text. It can be achieved by the method introduced in \cite{BAN20}, which generalizes diffusion maps to apply to SDEs of the form (\ref{eq:system}). In addition to the isotropic Kernel (\ref{eq:kernel}), an anisotropic Kernel with a second bandwidth $\epsilon_l$, as introduced in \cite{BER16}, is used: 
\begin{equation}
\label{eq:local_kernel}
K_{\epsilon_l}(\mathbf{x}, \mathbf{y}) = \exp \left(- (4 \epsilon_l)^{-1} \left(\mathbf{x} - \mathbf{y} + \epsilon_l \hat{b}(\mathbf{x})\right)^\mathsf{T} \left(\hat{A}(\mathbf{x})\right)^{-1} \left(\mathbf{x} - \mathbf{y} + \epsilon_l \hat{b}(\mathbf{x})\right) \right).
\end{equation}
This requires estimates of the drift $\hat{b}(\mathbf{x})$ and diffusion matrix $\hat{A}(\mathbf{x})$ locally at each data point $\mathbf{x}$. To do so we use Kramers-Moyal expansion estimates 
\begin{align}
 \hat{b}(\mathbf{x}) &= \lim_{\tau \to 0} \frac{1}{\tau} \mathbf{E} [X_{\tau} - X_0 | X_0 = \mathbf{x}] \\
 \hat{A}(\mathbf{x}) &= \lim_{\tau \to 0} \frac{1}{2 \tau} \mathbf{Cov} [X_{\tau} - X_0 | X_0 = \mathbf{x}]. 
\end{align}
Ideally, in order to calculate expectation values, one has many realizations with short integration time $\tau$ at each data point as initial condition $\mathbf{x}$. But in the case of the Veros model, we only have one long realization (for each parameter value), and thus we estimate drift and diffusion at each data point $\mathbf{x}$ by computing expectation and covariance over the set of increments $\{Y_{n_i +  \tau} - Y_{n_i}\}_{i=1}^{n}$, where $\{Y_{n_i}\}_{i=1}^{n}$ are the $n$ nearest neighbors of $\mathbf{x}$. While we cannot take the limit $\tau \to 0$, the sample time $\tau$ of 5 years is short compared to the long time scales of the deep ocean. 

After computing the anisotropic Kernel $(K_{\epsilon_l})_{ij} = K(\mathbf{x}_i, \mathbf{x}_j)$ for all pairs of data points, it is normalized using the row sum of the isotropic Kernel $p_{\epsilon}(\mathbf{x}_i) = \sum_{j=1}^{N} K(\mathbf{x}_i, \mathbf{x}_j)$, i.e., a Kernel density estimate of the data sample. Thus, with the diagonal matrix $(D_{\epsilon} )_{ii} = p_{\epsilon}^{-1}(\mathbf{x}_i)$, form $\tilde{K}_{\epsilon_l} = K_{\epsilon_l} D_{\epsilon}$. 
This normalization removes additional drift terms in the reconstructed operator, which are induced by the non-uniform sampling density. Finally, to conserve probability, the matrix is normalized by its row sums, i.e., $L_{\epsilon_l} = D_{\epsilon, \epsilon_l}^{-1} \tilde{K}_{\epsilon_l}$ with the diagonal matrix $(D_{\epsilon, \epsilon_l})_{ii} = \sum_{j=1}^{N} (\tilde{K}_{\epsilon_l})_{ij}$. As shown in \cite{BAN20}, in the limits $N \to \infty$, $\epsilon \to 0$, and $\epsilon_l \to 0$ the matrix  $(L_{\epsilon_l}-I)/\epsilon_l$ converges at every point to the generator (backward FP operator) $\mathcal{L}^*$ of the general SDE (\ref{eq:system}), and we can approximate $\{\phi_n\}$ by the eigenfunctions of $L_{\epsilon_l}$. The forward FP operator can be constructed in a similar fashion \cite{BAN20}. 

\begin{figure}%[floatfix]%!htb
\includegraphics[width=0.65\textwidth]{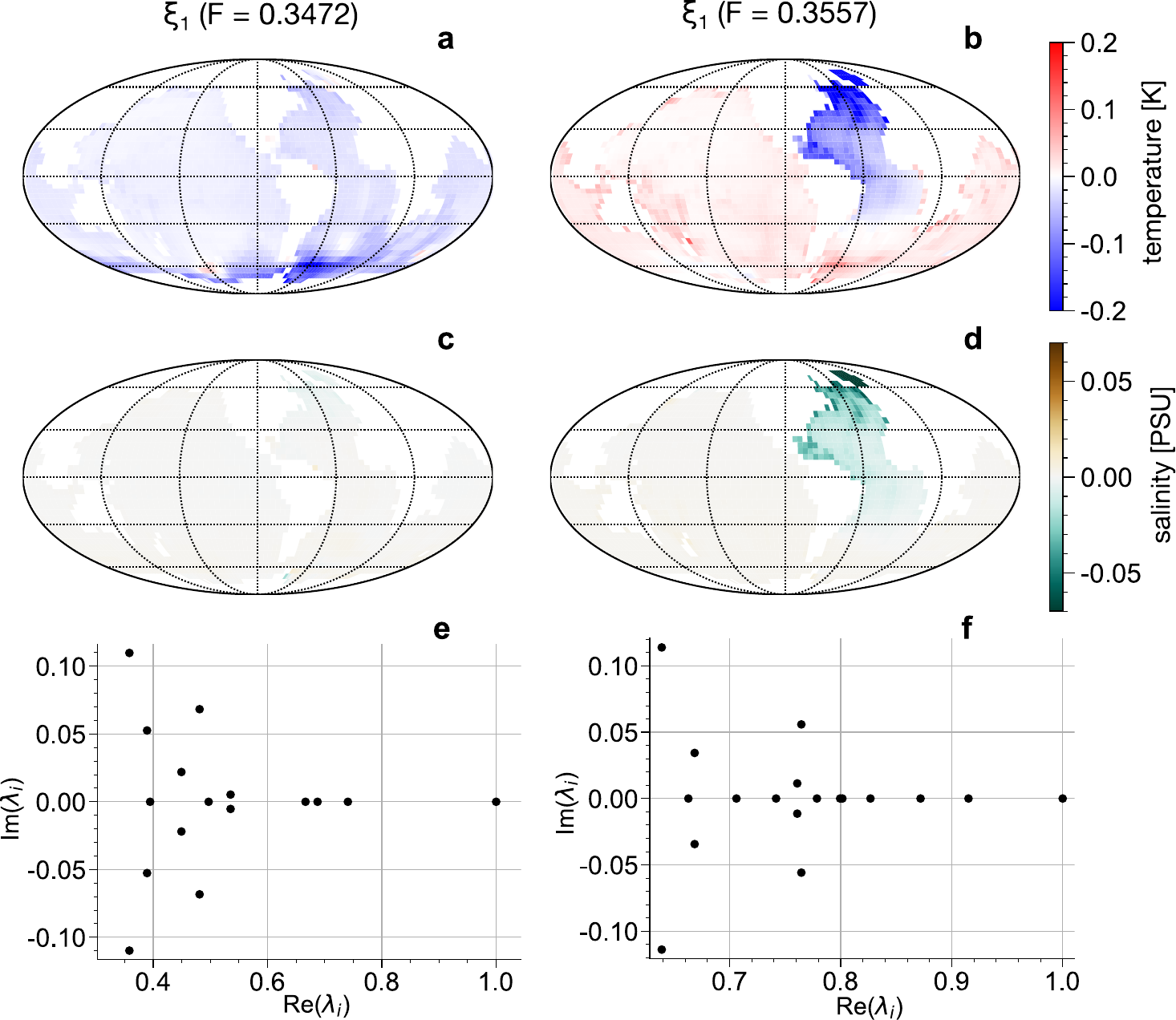}
\caption{\label{fig:veros_modes_localK} 
{\bf a-d} Anomaly maps of deep ocean (vertical average below 1000m) temperature ({\bf a,b}) and salinity ({\bf c,d}) describing the first subdominant spatiotemporal mode $\xi_1$ of the Veros model obtained by the local Kernel DM algorithm \cite{BAN20} at the control parameter values $F=0.3472$ ({\bf a,c}) and $F=0.3557$ ({\bf b,d}). The modes are constructed in the same way as described in Sec.~\ref{sec:veros}. 
{\bf e,f} Spectrum of the local Kernel diffusion map (higher eigenvalues discarded). Note that the reported eigenvalues of the matrix $L_{\epsilon_l}$ correspond to $e^{\lambda_n}$ of the approximated generator. 
}
\end{figure}

For the analysis of the Veros data, we chose $\epsilon_l / \tau = 5$ and $n=100$. As explained in \cite{BAN20}, the bandwidth may be interpreted as a characteristic time scale, and is chosen here to match the average displacement in phase space after one sample time $\tau$. The estimate of the inverse diffusion matrix is regularized by replacing $\hat{A}(\mathbf{x})$ with $\hat{A}(\mathbf{x}) + \eta I$ in (\ref{eq:local_kernel}). A large regularization parameter of $\eta = 0.3$ had to be chosen due to the small sample size. 
% this may make the noise close to isotropic and state-independent.
% No big deal here since noise not state-dependent (although quite anisotropic) 
% matching the large statistical uncertainty 
In Fig.~\ref{fig:veros_modes_localK} the results are summarized. We focus on the parameter very close to the bifurcation ($F=0.3557$), as well as one further away ($F=0.3472$), just as presented in Fig.~\ref{fig:veros_modes}. The evolution of the eigenfunctions is very similar to the isotropic diffusion map method. Since the generalized method also picks up non-gradient dynamics, there are oscillatory modes, but they relax more quickly and thus appear higher up in the spectrum (Fig.~\ref{fig:veros_modes_localK}e,f). The initially dominant mode of cold excursions initiated in the Southern ocean is picked up more clearly with the generalized method (Fig.~\ref{fig:veros_modes_localK}a,c). The critical mode close to the TP (Fig.~\ref{fig:veros_modes_localK}b,d) is again very similar to the fingerprint of the edge state (Fig.~\ref{fig:veros_modes}c,f) and thus allows strong EWS by projecting time series data onto it.

\section{Equations for the five-box ocean model}
\label{AppB}

In this appendix, the equation and parameter values of the five-box ocean model, originally published in \cite{WOO19}, are described. The boxes, labelled by $X={N,S,T,IP,B}$, are coupled unidirectionally by the thermohaline overturning circulation $q$, and bi-directionally by the wind-driven circulation. 
The dynamical equation for box $B$ can be eliminated by salt conservation. The remaining boxes forced by an atmospheric freshwater flux $F_X$ multiplied by the reference salinity $S_0 = 0.035$, which is then modulated by $H A_X$ to emulate the effect of climate change, where $H$ is the control parameter. The varying strength of the overturning $q$ is proportional to the density difference in the northern and southern boxes, and the temperatures $T_X$ are fixed everywhere except in the northern box, where it is assumed that $T_N = \mu q + T_0$, with a global reference temperature $T_0$.
This yields
\begin{equation}
    q =  \lambda\frac{  \alpha (T_{\rm{S}} - T_{\rm{0}}) + \beta (S_{\rm{N}} - S_{\rm{S}})}{1 + \lambda \alpha \mu }.
    \label{eq:q_2}
\end{equation}

In the model, $q>0$ corresponds to an AMOC `ON' state, and it is assumed that in case of a reversed circulation $q<0$ the unidirectional coupling by the overturning flow is reversed. This yields different dynamics for positive and negative $q$, and a non-smooth system of four ODEs, using the Heaviside function $\Theta (\cdot)$:
\begin{subequations}  \label{eq:5box}
\begin{flalign}
   & V_N \frac{dS_N}{dt} =  |q| \left[ \Theta(q) (S_T - S_B) + S_B - S_N \right] + K_N(S_T - S_N) - (F_N + H A_N) S_0 \\
   & V_T \frac{dS_T}{dt} = |q| \left[ \Theta(q) (\gamma S_S + (1-\gamma) S_{IP} -S_N) +S_N -S_T \right] + K_S(S_S - S_T)+ \notag \\ 
     & \qquad \qquad \qquad K_N(S_N - S_T) - (F_T + H A_T)  S_0  \\
    &V_S \frac{dS_S}{dt} =  \gamma |q| \left[\Theta(q)(S_B - S_T) + S_T - S_S \right] + K_{IP}(S_{IP} - S_S) + K_S(S_T - S_S) + \notag \\    
     & \qquad \qquad \qquad  \eta (S_B - S_S) - (F_S + H A_S) S_0 \\
    &V_{IP} \frac{dS_{IP}}{dt} = (1-\gamma) |q| \left[ \Theta(q) (S_B - S_T) + S_T - S_{IP} \right] + K_{IP}(S_S-S_{IP}) - (F_{IP} + H A_{IP}) S_0. 
   %& V_B \frac{dS_B}{dt} = |q| \left[ \Theta(q) (S_N - S_B - \gamma S_S - (1-\gamma) S_{IP}) + \gamma S_S + (1-\gamma)S_{IP} - S_B \right]   + \eta (S_S - S_B).
\end{flalign}
\end{subequations}

Time is re-scaled by $\tau_Y = 3.15 \times 10^{7}$ to go from seconds to years, and the remaining parameter values are listed in Tab.~\ref{tab:params}. Additive noise is included to yield stochastic differential equations of the form
\begin{equation}
    dS_X = f_X (S_X,H) dt + \sigma_X dW_X,
    \label{eq:SDE}
\end{equation}
with $X \in \{ N,T,S,IP \}$, $\sigma_X = 10^{-6}$. The drift $f_X$ represents the deterministic model (\ref{eq:5box}) and $W_X$ are standard independent Wiener processes.

\begingroup
\renewcommand{\arraystretch}{0.8} % Default value: 1
\begin{table}[h]
  \caption{Parameter values used for the five-box model, adapted from the {\tt FAMOUSA1xCO2} calibration in \cite{WOO19}. $\alpha = 0.12$ (thermal coefficient) and $\beta = 790$ (haline coefficient) define a linear equation of state for the density of sea water. $V_i$ is box volume,  $F_i$ the freshwater fluxes, $T$ are temperatures, $K_i$ are wind fluxes and $A_i$ determine the distribution of freshwater forcing. $\eta$ is a mixing parameter between the S and B boxes, $\gamma$ determines the proportion of water which takes the cold-water path, $\lambda$ and $\mu$ are constants. Subscripts indicate box labels, $i \in \{ N, T, S, IP, B \}$, and '$0$' indicates a global reference value.}
\label{tab:params}
\centering
%\begin{tabular}{| l || c |}
\begin{tabular}{ l | c || l | c }
%\hline
Parameter  & Value & Parameter & Value\\
\hline
\hline
$V_{\rm{N}} (m^3)$ &  3.683 $\times 10^{16}$ & $F_{\rm{N}} (m^3  s^{-1})$ &  0.375 $\times 10^{6}$\\ 
$V_{\rm{T}} (m^3)$  & 5.151 $\times 10^{16}$ & $F_{\rm{T}} (m^3  s^{-1})$ &  -0.723 $\times 10^{6}$\\ 
$V_{\rm{S}} (m^3)$ &  10.28 $\times 10^{16}$ & $F_{\rm{S}} (m^3  s^{-1})$ &  1.014 $\times 10^{6}$\\
$V_{\rm{IP}} (m^3)$  & 21.29 $\times 10^{16}$ & $F_{\rm{IP}} (m^3  s^{-1})$ &  -0.666 $\times 10^{6}$\\
$V_{\rm{B}} (m^3)$  & 88.12 $\times 10^{16}$ &  $F_{\rm{B}} (m^3  s^{-1})$  &  0  \\ 
\hline
$A_{\rm{N}}$  & 0.194 & $\eta (m^3  s^{-1})$ & 66.061 $\times 10^{6}$\\
$A_{\rm{T}}$ & 0.597 & $\gamma$  & 0.1 \\
$A_{\rm{S}}$ & -0.226 & $\lambda (m^6 kg^{-1} s^{-1})$ & 2.66 $\times 10^{7}$\\
$A_{\rm{IP}}$  & -0.565 & $\mu (^o C m^{-3}s)$ & 7.0 $\times 10^{-8}$\\ 
\hline
$K_{\rm{N}} (m^3  s^{-1})$ & 5.439 $\times 10^{6}$& $T_{\rm{S}} (^o C)$ &  5.571 \\
$K_{\rm{S}} (m^3  s^{-1})$  & 3.760 $\times 10^{6}$& $T_{\rm{0}} (^o C)$  & 3.26 \\ 
$K_{\rm{IP}} (m^3  s^{-1})$  & 89.778 $\times 10^{6}$& & \\
%\hline
\end{tabular}
\end{table}
\endgroup

\end{document}